\documentstyle[aps,twocolumn]{revtex}
\input epsf
\tightenlines
\begin{document}
\draft
\preprint{DUKE-TH-96-126}
\title{\bf Expanding Quark-Gluon Plasmas: Transverse Flow, Chemical
Equilibration and Electromagnetic Radiation}
\author{Dinesh Kumar Srivastava, Munshi Golam Mustafa}
\address{Variable Energy Cyclotron center, 1/AF Bidhan Nagar, Calcutta 
700 064}
\author{Berndt M\"{u}ller}
\address{Department of Physics, Duke University, Durham, North Carolina
27708-0305}
\date{\today}
\maketitle

\begin{abstract}
We investigate the chemical equilibration of the parton distributions 
in collisions of two heavy nuclei.  We use initial conditions obtained 
from a self-screened parton cascade calculation and, for comparison,
from the HIJING model.  We consider a one-dimensional, as well as a 
three-dimensional expansion of the parton plasma and find that the 
onset of the transverse expansion impedes the chemical equilibration. 
At energies of 100 GeV/nucleon, the results for one-dimensional and 
three-dimensional expansion are quite similar except at large values 
of the transverse radius.  At energies of several TeV/nucleon, the 
plasma initially approaches chemical equilibrium, but then is 
driven away from it, when the transverse velocity  gradients develop. 
We find that the total parton multiplicity density remains essentially 
unaffected by the flow, but the individual concentrations of quarks, 
antiquarks, and gluons are sensitive to the transverse flow.  
The consequences of the flow are also discernible in the transverse
momenta of the partons and in the lepton pair spectra, where the flow
causes a violation of the so-called $M_T$ scaling.

\end{abstract}
%
%
\narrowtext

\section{\bf INTRODUCTION}

Relativistic heavy ion collisions are being studied with the intention of 
investigating the properties of ultradense, strongly interacting matter.
A central theme of this area of research is the search for a quark-gluon 
plasma (QGP) \cite{HM96}.  Extensive theoretical studies done over the 
last several years for nuclear reactions at collider energies have helped 
develop a picture of the colliding nuclei as two clouds of valence and 
sea partons which interact passing through each other \cite{klaus}. 
Partonic final-state interactions are then thought to produce a dense plasma
of quarks and gluons.  This plasma expands and becomes cooler and more dilute. 
If quantum chromodynamics admits a first-order deconfinement or chiral 
phase transition, this plasma will pass through a mixed phase of quarks, 
gluons, and hadrons, before the hadrons lose thermal contact and stream 
freely towards the detectors.  

What is the structure of the matter formed in these nuclear collisions?  
Does the initial partonic system attain kinetic equilibrium? 
Probably yes, as the initial parton density is 
large, forcing the partons to suffer many collisions in a very short 
time \cite{kin}.  Does it attain chemical equilibration?  This will 
depend on the time available \cite{biro,chem,kg,ja,xmh} to the partonic system 
before it converts into a mixed phase, or before perturbative
QCD is no longer applicable.
The time available for equilibration is perhaps  too short (3--5 fm/c) at the
energies ($\sqrt{s} \leq 100$ GeV/nucleon) accessible to the 
Relativistic Heavy Ion Collider (RHIC). At the energies ($\sqrt{s} 
\leq 3$ TeV/nucleon) that will be achieved at the CERN Large Hadron
Collider (LHC) this time could be large (more than 10 fm/$c$).
If one considered only a longitudinal expansion of the system, 
the QGP formed at LHC energies could approach chemical equilibrium 
very closely, due to the higher initial temperature predicted to be 
attained there.  However, then the life-time of the plasma would also be 
large enough to allow a rarefaction wave from the surface of the plasma 
to propagate to the center. As we shall discuss below, the large transverse 
velocity gradients developing in this process at LHC  energies 
may drive the system away from the chemical equilibrium \cite{munshi}.

It is evident that the evolution of the 
partonic system will depend on the initial conditions and the total
time available to the system where the partonic picture still holds.
In order to study this aspect, we use initial conditions from two
models: the Self-Screened Parton Cascade (SSPC) model \cite{sspc} and the
HIJING model \cite{hijing}.

By now, a considerable progress has been achieved in our understanding of 
the parton cascades which develop in the wake of the collisions. 
Early calculations \cite{klaus} were done by assuming fixed $p_T$- and 
virtuality cut-offs for the partonic interactions to ensure the 
applicability of the perturbative expansion for QCD scattering processes.  
In the recently formulated self-screened parton
cascade model early hard scatterings produce a medium which screens the longer
ranged color fields associated with softer interactions. When two
heavy nuclei collide at sufficiently high energy, the screening occurs
on a length scale where perturbative QCD still applies.
This approach yields predictions for the initial conditions of the
forming QGP without the need for any {\it ad-hoc} momentum and virtuality
cut-off parameters. These calculations also show that the QGP
likely to be formed in such collisions could be very hot and initially
far from chemical equilibrium. 

It is still useful to explore the dependence of  the course of evolution
of the plasma on the initial conditions.
For this purpose we use the results of the HIJING  Monte Carlo model
which combines multiple hard or semihard parton scatterings with initial-
and final-state radiation with Lund string phenomenology
for the accompanying soft nonperturbative interactions to describe
nuclear interactions.
The uncertainties  of the model are parametrized by allowing for
variations of the initial fugacities. This approach has, for example, been
used recently \cite{xu} in connection with the suppression of
 J/$\psi$ in an equilibrating partonic plasma.

In the next section we briefly describe the hydrodynamic and chemical
evolution of the plasma in a (1+1) dimensional longitudinal expansion
and a (3+1) dimensional transverse expansion.
The thermal photon and thermal lepton pair production is 
discussed in Section III, followed by a summary in Section IV. 

\section{\bf HYDRODYNAMIC EXPANSION AND CHEMICAL EQUILIBRATION}
\subsection{Basic Equations}

We start with the assumption that the system achieves kinetic equilibrium
by the time $\tau_i$. We define the beginning of this epoch as the instant
beyond which the momenta of the partons become locally isotropic. 
Local isotropy here is defined as the coincidence of the exponential
slopes of the longitudinal and transverse momentum distributions, when
they are integrated over a comoving volume with linear dimensions of
one partonic mean free path $\lambda_f$.  This occurs after a proper 
time of about $0.7\,\lambda_f$ \cite{kin}, corresponding to 
$\tau_i\approx 0.2-0.3$ fm/$c$ under the conditions expected at 
collider energies \cite{biro}.  Beyond this point,
further expansion can be described by hydrodynamic equations. 
The approach to the chemical equilibration is then governed by 
a set of master equations which are driven by the two-body reactions
($gg\,\leftrightarrow\,q\bar{q}$) and gluon multiplication and its
inverse process, gluon fusion ($gg\,\leftrightarrow\,ggg$).
The hot matter continues to expand and cools due to expansion and
chemical equilibration.  Once the energy density reaches a critical
value (here taken as $\epsilon_f=1.45$ GeV/fm$^3$ \cite{com}) we 
terminate the evolution before entering into the hadronization phase.

Thus the expansion of the system is described by the equation for 
conservation of energy and momentum of an ideal fluid:
\begin{equation}
\partial_\mu T^{\mu \nu}=0 \; , \qquad
 T^{\mu \nu}=(\epsilon+P) u^\mu u^\nu + P g^{\mu \nu} \, ,
\label{hydro}
\end{equation}
where $\epsilon$ is the energy density and $P$ is the pressure measured 
in the frame comoving with the fluid. 
The four-velocity vector $u^\mu$ of the 
fluid satisfies the constraint $u^2=-1$. For a partially equilibrated 
plasma of massless particles, the equation of state can be written as
\cite{biro}
\begin{equation}
\epsilon=3P=\left[a_2 \lambda_g +  b_2 \left (\lambda_q+\lambda_{\bar q}
\right ) \right] T^4 \, ,
\label{eos}
\end{equation}
where $a_2=8\pi^2/15$, $b_2=7\pi^2 N_f/40$, $N_f \approx 2.5$ is 
the number of dynamical quark flavors, and $\lambda_k$ is the fugacity
for the parton species $k$.  Here we  have defined the fugacities 
through the relations,
\begin{equation}
n_g=\lambda_g \tilde{n}_g,\qquad 
 n_q=\lambda_q \tilde{n}_q,
\end{equation}
where $\tilde{n}_k$ is the equilibrium density for the parton species $k$:
\begin{equation}
\tilde{n}_g=\frac{16}{\pi^2}\zeta(3) T^3=a_1 T^3,
\end{equation}
\begin{equation}
\tilde{n}_q=\frac{9}{2\pi^2}\zeta(3) N_f T^3=b_1 T^3.
\end{equation}
We further assume that $\lambda_q=\lambda_{\bar{q}}$.  The equation of
state (\ref{eos}) implies the speed of sound $c_s=1/\sqrt{3}$.

We solve the hydrodynamic equations (\ref{hydro}) with the assumption
that the system undergoes a boost invariant longitudinal expansion along
the $z$-axis and a cylindrically symmetric transverse expansion \cite{vesa}.
It is then sufficient to solve the problem for $z=0$, because of the 
assumption of boost invariance. 

The master equations \cite{biro} for the dominant chemical reactions 
$gg \leftrightarrow ggg$ and $gg \leftrightarrow q\bar{q}$  are
\begin{eqnarray}
\partial_\mu (n_g u^\mu)&=&n_g(R_{2 \rightarrow 3} -R_{3 \rightarrow 2})
                    - (n_g R_{g \rightarrow q}
                       -n_q R_{q \rightarrow g} ) \, , \nonumber\\
\partial_\mu (n_q u^\mu)&=&\partial_\mu (n_{\bar{q}} u^\mu)
                     = n_g R_{g \rightarrow q}
                       -n_q R_{q \rightarrow g},
\label{master1}
\end{eqnarray}
in an obvious notation. 

If we assume the system to undergo  a purely longitudinal boost invariant 
expansion, (\ref{hydro}) reduces to the well known relation \cite{bj}
\begin{equation}
\frac{d\epsilon}{d\tau}+\frac{\epsilon+P}{\tau}=0,
\label{long}
\end{equation}
where $\tau$ is the proper time. This equation implies
\begin{equation}
\epsilon\, \tau^{4/3}=\,{\rm {const}.}
\label{epstau}
\end{equation}
and the chemical master equations reduce to \cite{biro}
\begin{eqnarray}
\frac{1}{\lambda_g}\frac{d \lambda_g}{d\tau}
+\frac{3}{T}\frac{dT}{d\tau} +
\frac{1}{\tau} 
 &=&
R_3 ( 1- \lambda_g ) -2 R_2 \left( 1-\frac{\lambda_q \lambda_{\bar{q}}}
{\lambda_g^2}\right) \, , 
\nonumber\\
\frac{1}{\lambda_q}\frac{d \lambda_q}{d\tau}
+\frac{3}{T}\frac{dT}{d\tau} +
\frac{1}{\tau} 
 &=&
R_2 \frac{a_1}{b_1} \left(
\frac{\lambda_g}{\lambda_q}-\frac{\lambda_{\bar{q}}}{\lambda_g}\right)\, ,
\label{master_long}
\end{eqnarray}
which are then solved numerically for the fugacities. 
The rate constants $R_2$ and $R_3$ are related to the rates appearing in 
(\ref{master1}) and are given by \cite{biro}
\begin{eqnarray}
R_2 & \approx & 0.24 N_f \alpha_s^2 \lambda_g T \ln (1.65/\alpha_s \lambda_g),
\nonumber\\
R_3 & = & 1.2 \alpha_s^2 T (2\lambda_g-\lambda_g^2)^{1/2},
\end{eqnarray}
where the color Debye screening and the Landau-Pomeranchuk-Migdal effect
suppressing the induced gluon radiation have been taken into account,
explicitly.

In case of transverse expansion, the  master equations  can be shown
\cite{munshi} to lead to partial differential equations:
\begin{eqnarray}
\frac{\gamma}{\lambda_g}\partial_t \lambda_g &+& \frac{\gamma v_r }{\lambda_g}
\partial_r \lambda_g +\frac{1}{T^3}\partial_t (\gamma T^3) + \frac{v_r}{T^3}
\partial_r (\gamma T^3)  \nonumber\\ 
&+& \gamma \partial_r v_r +\gamma \left( \frac{v_r}{r}+\frac{1}{t}\right) 
\nonumber\\ &=& 
R_3 ( 1- \lambda_g ) -2 R_2 \left( 1-\frac{\lambda_q \lambda_{\bar{q}}}
{\lambda_g^2}\right) \, , 
\nonumber\\
\frac{\gamma}{\lambda_q}\partial_t \lambda_q &+& \frac{\gamma v_r }{\lambda_q}
\partial_r \lambda_q +\frac{1}{T^3}\partial_t (\gamma T^3) + \frac{v_r}{T^3}
\partial_r (\gamma T^3) \nonumber\\
&+& \gamma \partial_r v_r +\gamma \left( \frac{v_r}{r}+\frac{1}{t}\right) 
\nonumber\\ &=&
R_2 \frac{a_1}{b_1} \left(
\frac{\lambda_g}{\lambda_q}-\frac{\lambda_{\bar{q}}}{\lambda_g}\right)\, ,
\label{master}
\end{eqnarray}
where $v_r$ is the transverse velocity and $\gamma=1/\sqrt{1-v_r^2}$.
It is easy to verify that the equations (\ref{master}) reduce
to (\ref{master_long}) in the absence of transverse expansion.

The hydrodynamic equations (\ref{hydro}) are solved numerically to get
$\epsilon(r,t)$ and $v_r(r,t)$, which serve as input into the equations
(\ref{master}) for the fugacities.  In all our solutions we have assumed
that the initial transverse velocities are zero. 
We have verified that our results near $r=0$ closely follow the results 
for a purely longitudinal expansion, till the time when the fluid is 
disturbed by the rarefaction wave traveling from the surface to the center. 

\subsection{Results for RHIC Energies}

In the following, we shall discuss the results obtained for the initial 
conditions listed in Table I, which are predicted for a central collision 
of gold nuclei.  We shall discuss the results for the initial conditions 
from SSPC calculations in some detail, and then present the final results 
for the initial conditions from the HIJING model.

Let us first look at conditions likely to prevail at RHIC energies.
For the case of a boost-invariant and a purely longitudinal expansion,
the results are discussed in the literature \cite{new}.
Here we concentrate on the question, how the transverse expansion of 
the plasma affects the chemical equilibration, and how the consequences
of the transverse flow can best be identified.

Recall that for a boost-invariant longitudinal expansion, Eq.(\ref{hydro})
provides that $\epsilon \, \tau^{4/3}$ is a constant (\ref{epstau}).
Let us look at the constant energy density contours,
\begin{equation}
\epsilon(r,t)=\epsilon_i/N^{4/3}
\label{contour}
\end{equation}
for $N=1,2,3,\ldots,16$ (Figure 1a). 
The choice of hypersurfaces of equal comoving energy density is
natural, because it interpolates smoothly between the initial
condition and the final hadronization hypersurface, where the
quark-gluon plasma freezes out into hadrons. 
If there were no transverse expansion, all these contours would be parallel 
to the line for $N=1$ and given by
\begin{equation}
\tau=N\,\tau_i; \qquad ({\rm {longitudinal\,\, flow\,\, only}}),
\label{time}
\end{equation}
extending up to $r=R_T$.

For the initial energy density assumed at RHIC, $N=16$ corresponds to 
an energy density of about 1.5 GeV/fm$^3$, which coincides with our
choice of the density $\epsilon_f$ where the plasma finally hadronizes.
We have also shown a line $r=R_T-c_s t$ which indicates the radial size
of the region not yet affected by the rarefaction wave at any given time $t$.
We see that,  as expected, the fluid beyond $r=$ 4 fm is
likely to be affected by the flow, by the time the system has
cooled to the edge of the QGP phase.

In Fig.~1b--d we have given the gluon fugacity, the quark fugacity, and the
transverse velocity along these contours. We see that deviations from the
expectations for a longitudinal expansion \cite{new}, which we closely 
reproduce for small $r$, have a clear origin in the growth of the transverse
velocity. 

\begin{figure}
\epsfxsize=3.25in
\epsfbox{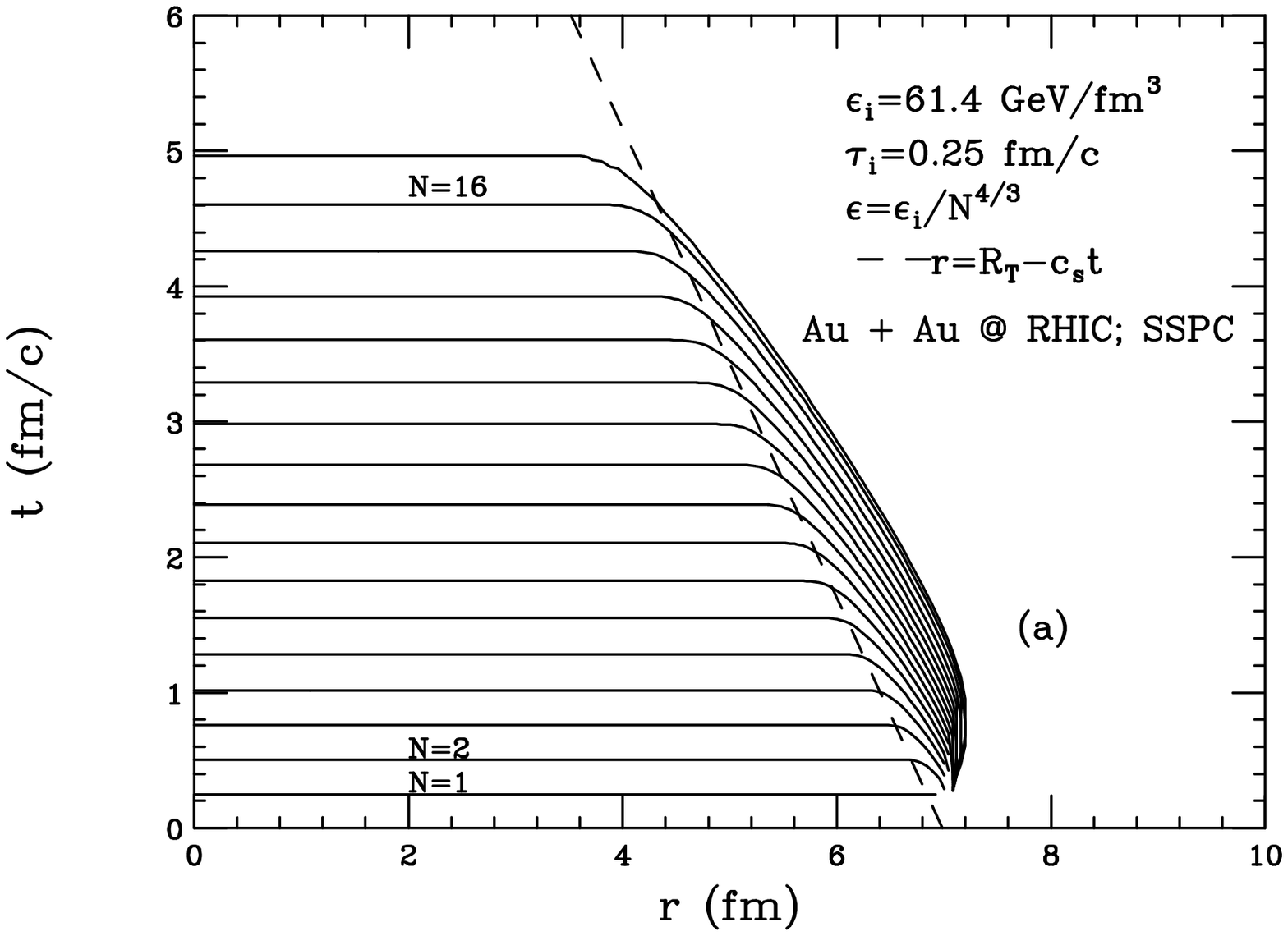}
\epsfxsize=3.25in
\epsfbox{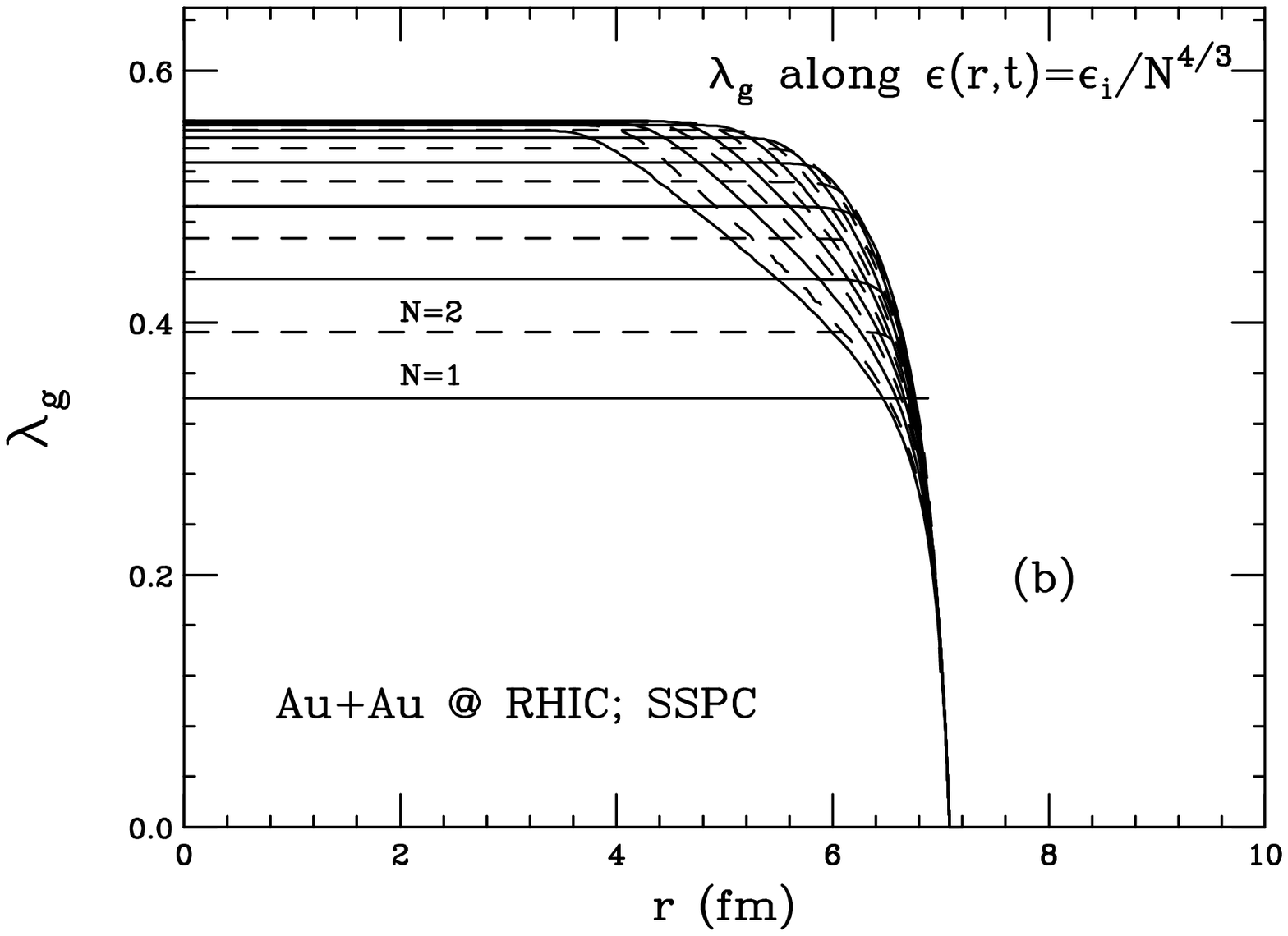}
\epsfxsize=3.25in
\epsfbox{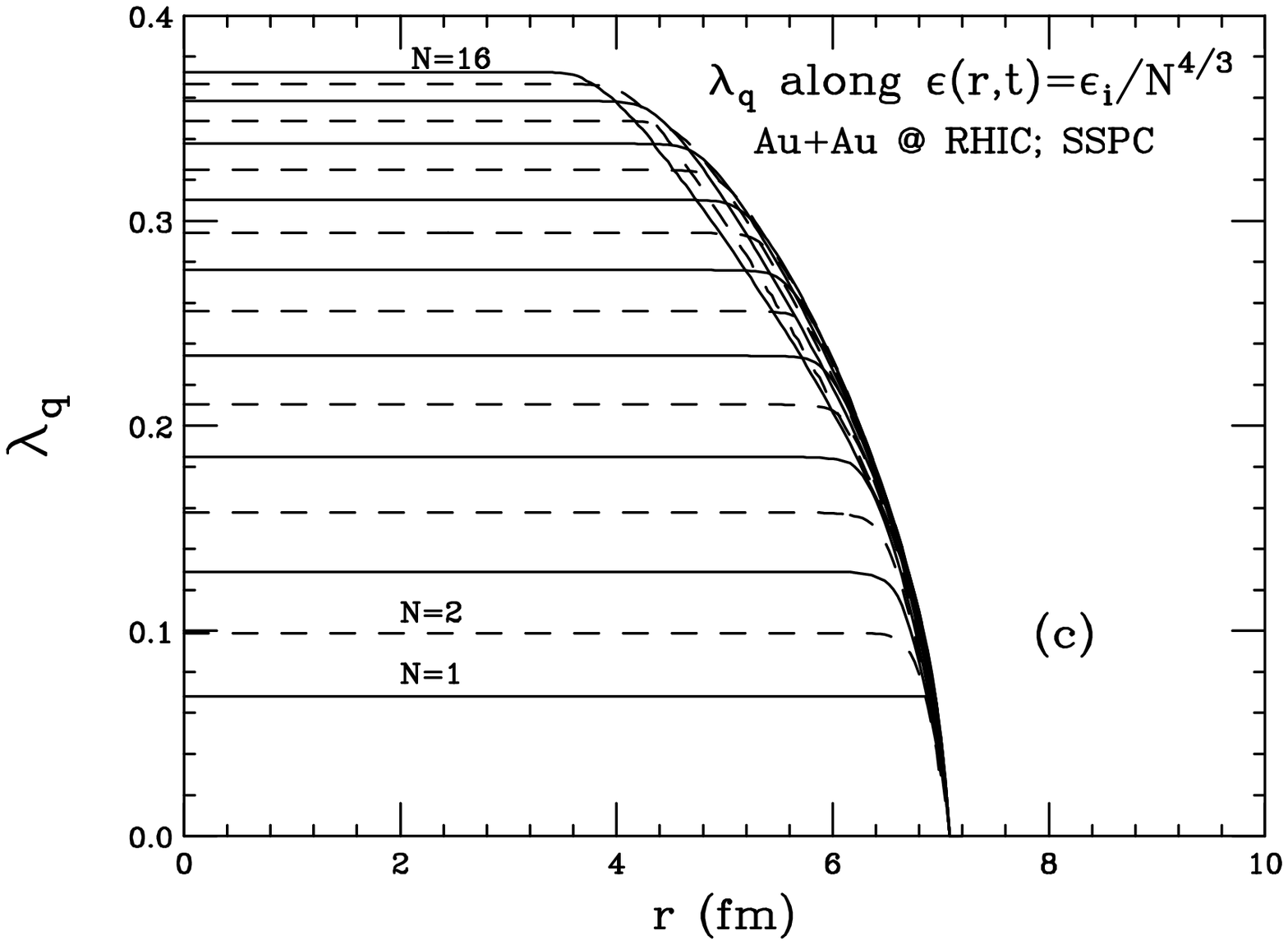}
\epsfxsize=3.25in
\epsfbox{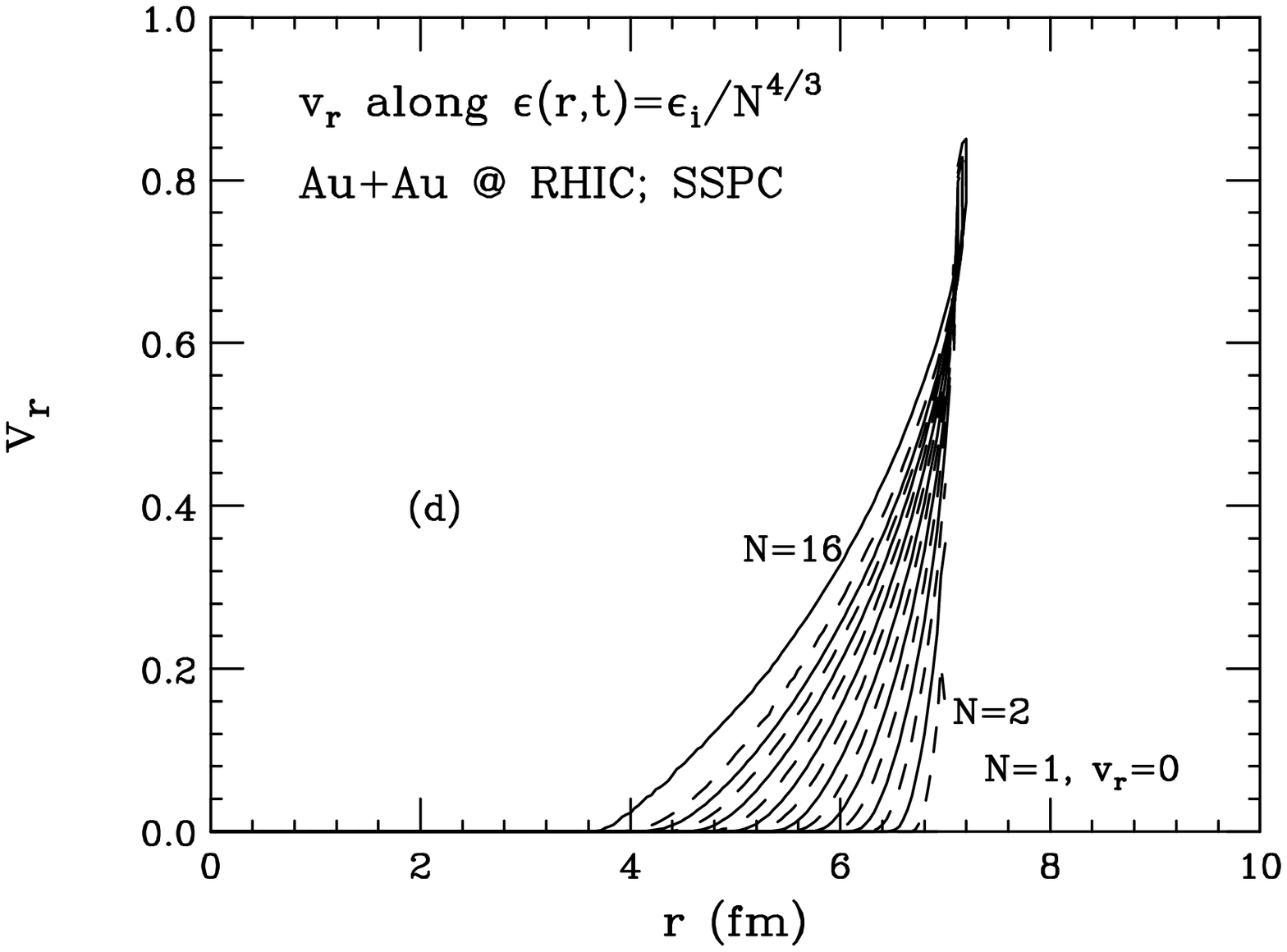}
\caption{(a) Constant energy density contours for a transversely
expanding quark-gluon plasma likely to be created in Au+Au collisions at
RHIC. For a boost invariant longitudinal expansion these contours would be
equidistant lines parallel to $N=1$. The 
dashed line gives the distance from the axis where the rarefaction
wave has arrived at time $t$, and beyond which the fluid is strongly 
affected by the flow.
(b) Gluon fugacity along the constant energy density contours.
(c) Quark fugacities along the constant energy density contours.
(d) Transverse velocities along the constant energy
density contours.}
\end{figure}

For small values of $N$, we see that the plasma at the center has more 
time at its disposal to develop while at large values of $r$ the time 
available to the plasma is shorter, leading to  a decrease in the 
fugacities (at constant $\epsilon$) as $r$ increases. 
However, this is not the most interesting observation.
We see a unique feature, which is quite evident at large $r$ but
is beginning to emerge at smaller $r$ as well, especially
for the gluon fugacity:  The fugacities initially increase  with $N$, 
but then start decreasing. Thus the region
where transverse velocities are large, the plasma  ends its journey
further away from chemical equilibrium, as compared to the region where the
transverse velocity is still small. We shall see later that this has its
origin in the velocity gradients which develop in the fluid.

This has an interesting consequence, namely, the plasma nearer to the 
surface will be hotter than in the interior, for a given energy density.
If we make the reasonable assumption that the transition point between
the quark-gluon plasma and the hadron phase is determined by a certain
critical value of the energy density, then our result implies that the
outer regions of the cylindrical reaction zone hadronize at a higher
temperature than the inner region. In other words, it is quite likely
that the hadronic matter may be formed at different temperatures in
different regions.  This enhances the complexity of the final state
produced in the hydrodynamic model compared to what was expected earlier. 

How does the flow affect the momentum distribution of the partons?
We can evaluate the evolution of the parton distributions with time
for the longitudinal expansion, and compare it with the corresponding
results for the transverse expansion, using the Cooper-Frye formula 
\cite{cf,acta}; 
\begin{equation}
\frac{dN}{d^3\vec{p}/E}=\frac{dN}{d^2p_T\,dy}=
\frac{g}{(2\pi)^3}\int_\sigma f(x,p) p^\mu d\sigma_\mu,
\end{equation}
where $f(x,p)$ is the phase-space distribution of the parton, $g$ is the
color, spin, and flavor degeneracy, and $\sigma$ is the surface 
described by the contours given by (\ref{contour}) or (\ref{time}). 
For massless particles and in the absence of transverse flow simplifications 
arise, which we shall not discuss here \cite{acta}.

The resulting transverse momentum spectra for gluons and quarks is shown
in Figs.~2a--b. We see that the parton distributions at large $p_T$
are affected strongly  as the flow develops. It is known that if the plasma
were to undergo a first order phase transition, then the nonvanishing
transverse velocity at the beginning of the mixed phase will reduce
its life-time.  We shall see, however, that due to several other competing 
factors, like lower temperature at later times, the flow only marginally
affects the production of thermal photons and lepton pairs from the 
quark-gluon phase at RHIC energies.  The presence of significant 
collective transverse flow is clearly visible in the positive curvature
of the $p_T$-spectra at late times ($N=12,16$).  The slope of these spectra 
at large $p_T$, resulting from a superposition of thermal and collective
motion, agrees remarkably well with the homogeneous slope of the
initial spectrum ($N=1$) and thus serves as an indicator of the very
high  apparent temperature achieved at the moment of thermalization.

\begin{figure}
\epsfxsize=3.25in
\epsfbox{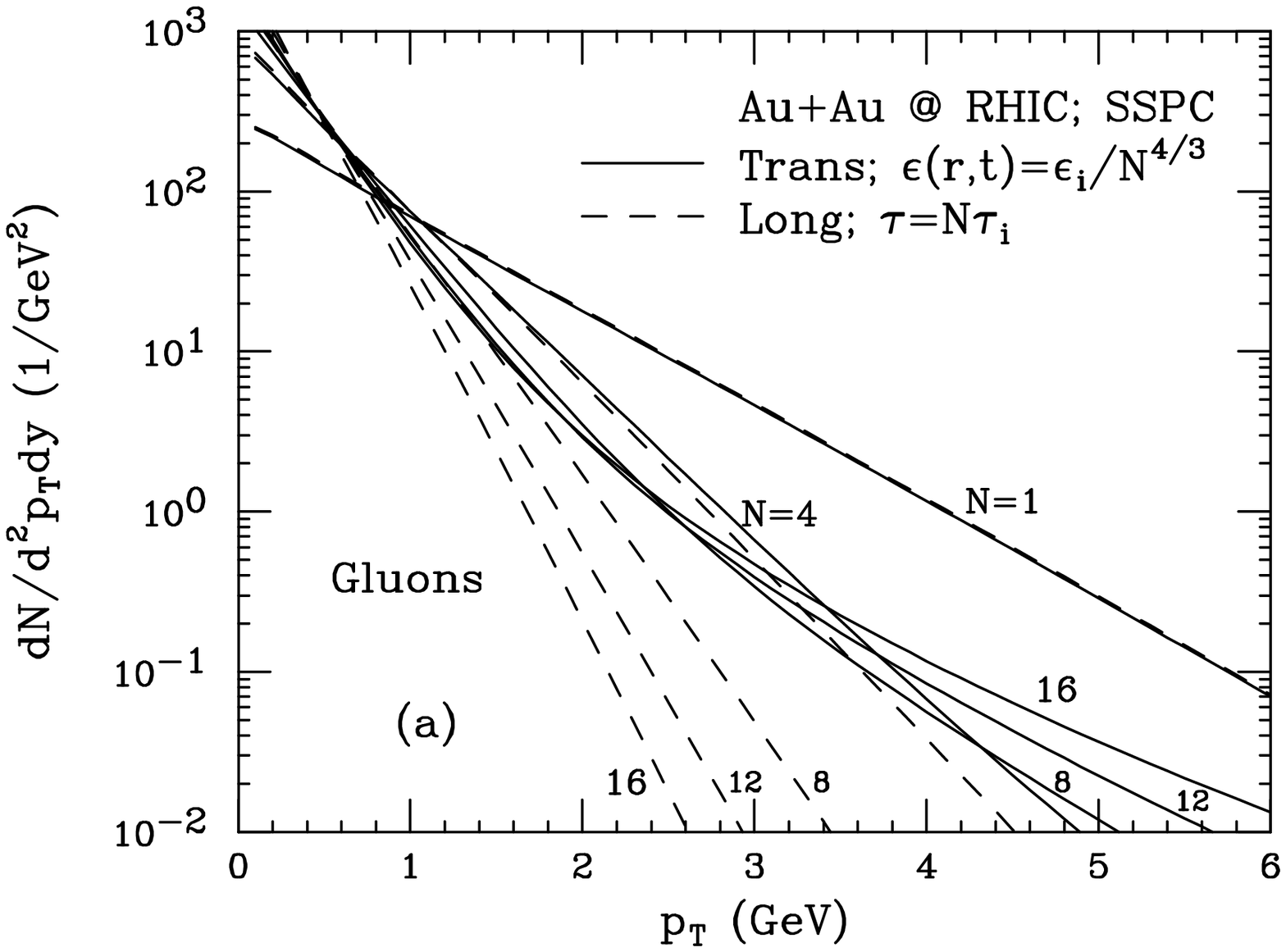}
\epsfxsize=3.25in
\epsfbox{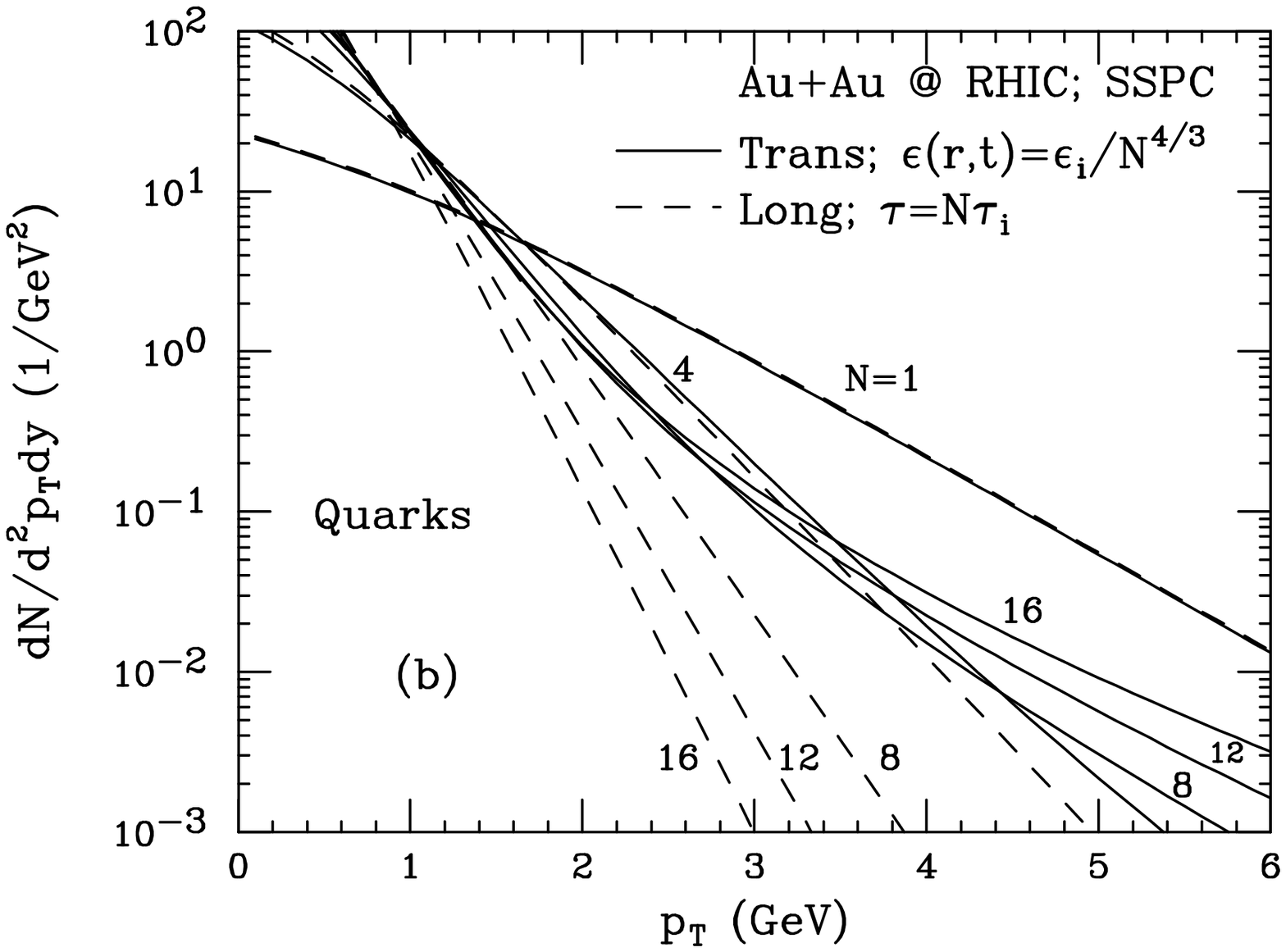}
\epsfxsize=3.25in
\epsfbox{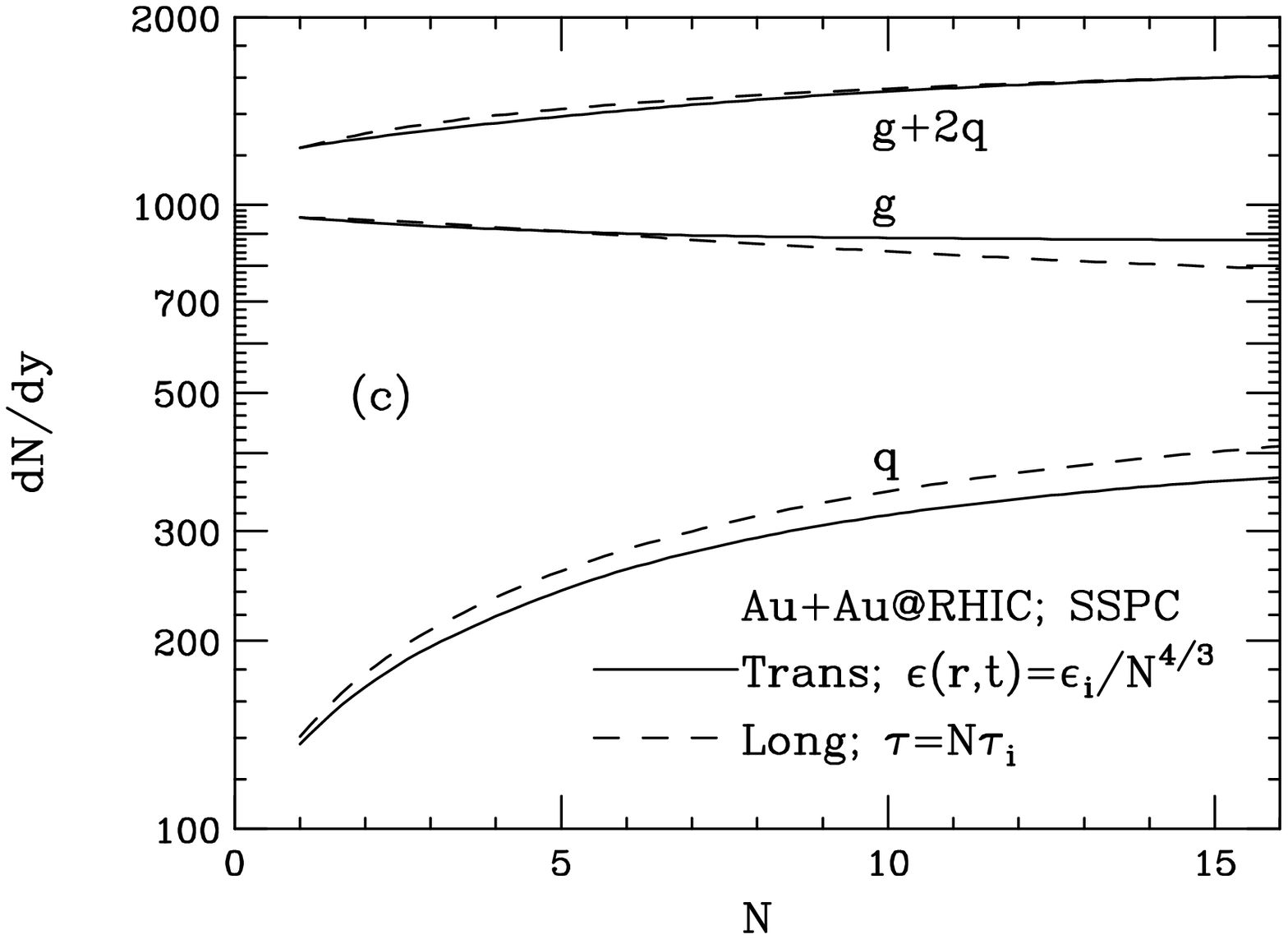}
\caption{ Evolution of the momentum distributions of (a) gluons and 
(b) quarks, and (c) of the multiplicity of partons 
with (solid curves) and without (dashed curves) transverse flow at RHIC.}
\end{figure}

We can also evaluate the multiplicity density of the partons as a function
of $N$ to see the increase in entropy, as the chemical equilibration proceeds,
and investigate its dependence on the flow. We see from Fig.~2c that 
the total entropy at the end of the QGP phase is essentially identical
for the scenarios with and without transverse flow. We also see that as the
chemical equilibration is accompanied by a steady increase in the net 
number of quarks (and antiquarks) whereas the number of gluons decreases.

However, the detailed composition of the partonic matter at the end of the
QGP phase is different for the case with flow. We have approximately 12\% 
less quarks and antiquarks but about 10\% more gluons compared to our 
result without flow. 
It is difficult to predict how these rather subtle differences in the
composition of the plasma will affect the final particle yields.
We may speculate that an increased gluon abundance could result in
the enhanced production of hadrons with a significant ``valence-glue''
component, such as $\eta$- and $\eta'$-mesons or even glueballs.

\subsection{Results for LHC Energies}

Let us next turn to the results for Au+Au collisions at LHC energies,
some of which have been reported earlier \cite{munshi}.
In Figs.~3a--d, we present the constant energy density contours, the 
fugacities for gluons and quarks, and the transverse velocities. The
representation is the same as the one for the RHIC-related figures,
except that we have plotted only a reduced number of equal-energy
contours.  Due to the very large number of contours between the initial
energy density and the hadronization density $\epsilon_f=1.45$ GeV/fm$^3$
we have plotted only the contours $N= 1,6,11,\ldots,71$ in the 
energy density plot and $N=1,11,21,\ldots,71$ in the plots for the 
other three quantities.

\begin{figure}
\epsfxsize=3.25in
\epsfbox{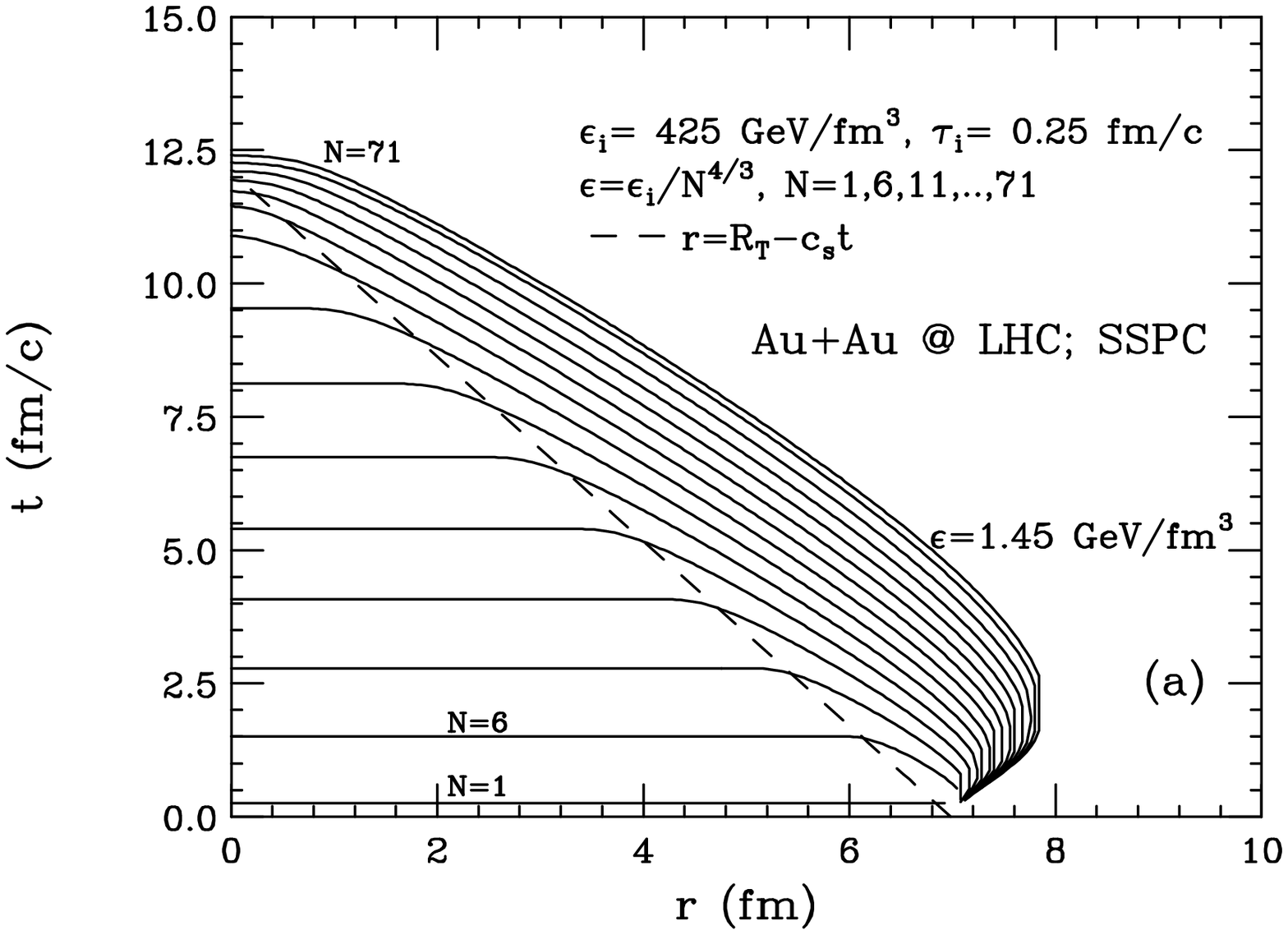}
\epsfxsize=3.25in
\epsfbox{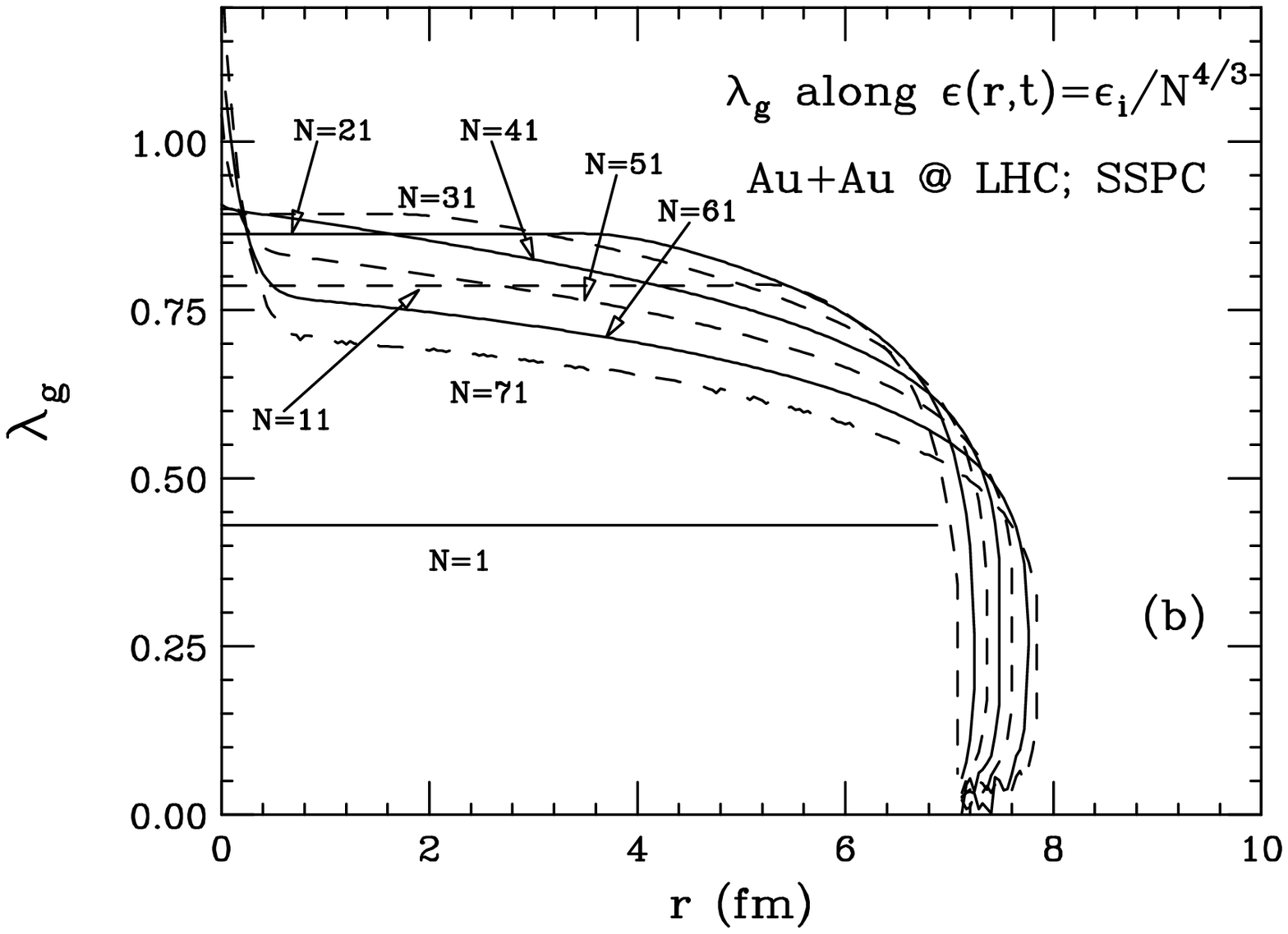}
\epsfxsize=3.25in
\epsfbox{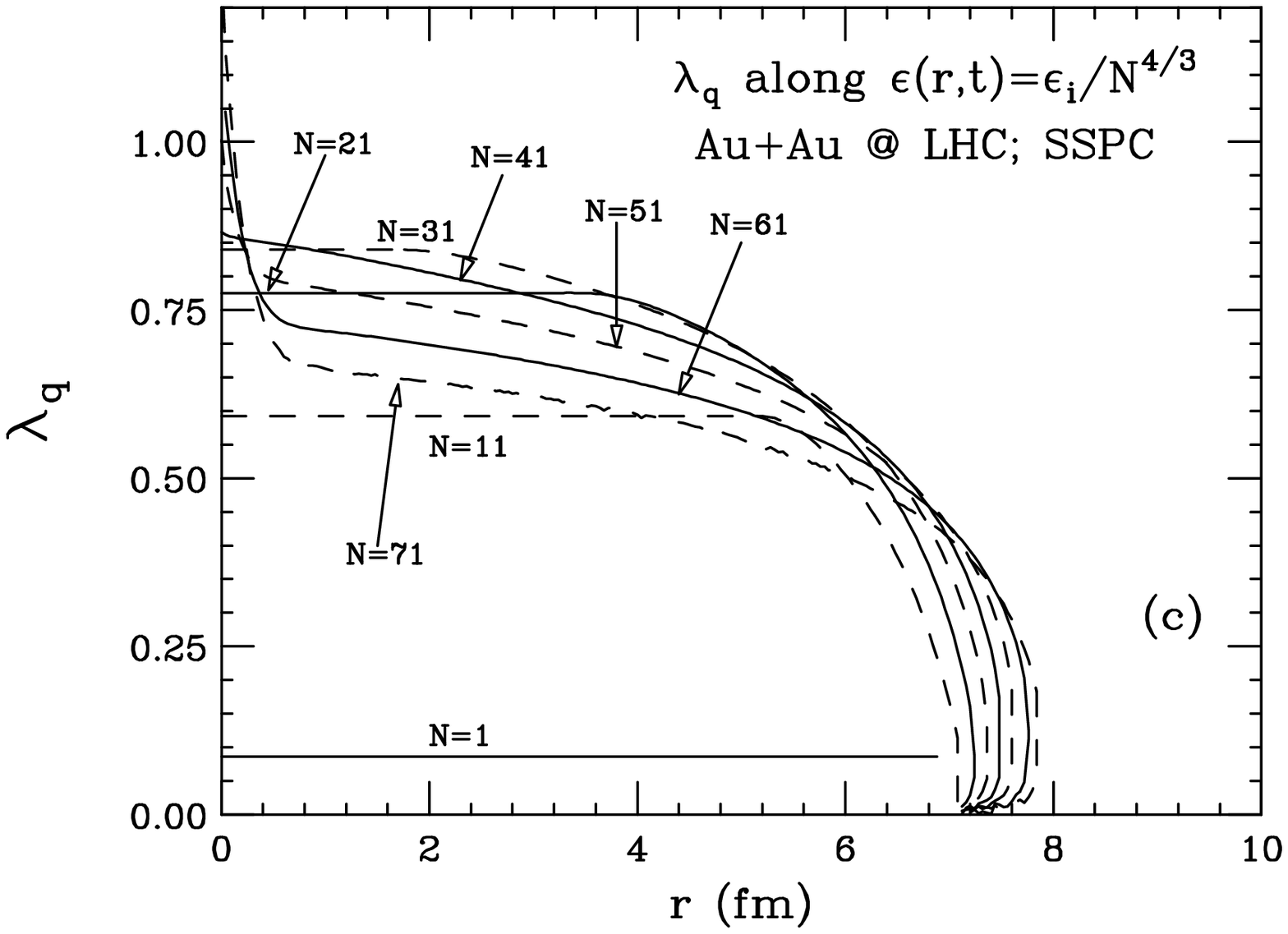}
\epsfxsize=3.25in
\epsfbox{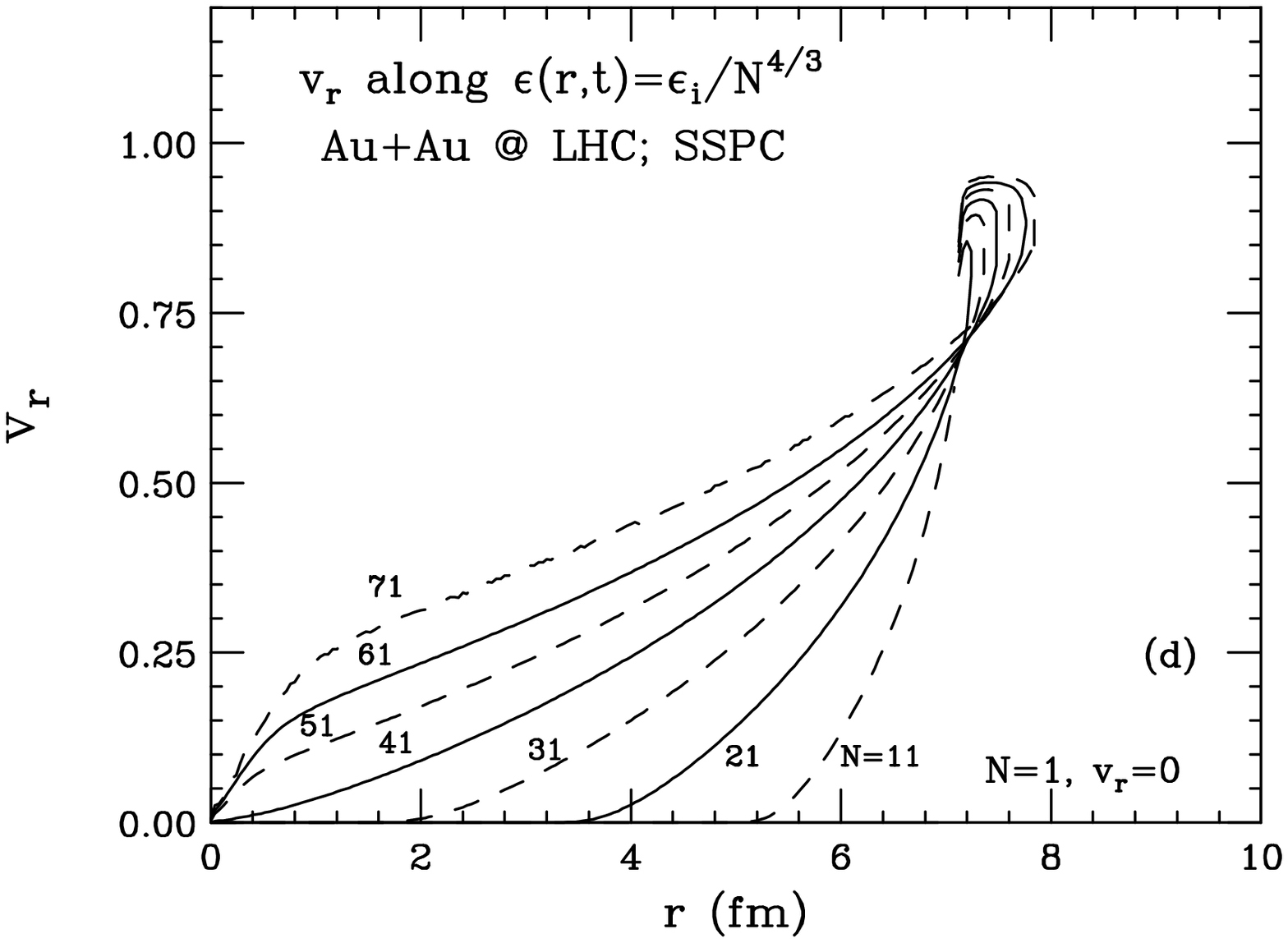}
\caption{(a)--(d): Same as Fig.~1, at LHC.}
\end{figure}

The same effects pertaining to the collective transverse flow
are observed as at the lower RHIC energy, but they are more pronounced.
Due to the longer lifetime of the plasma phase at LHC energies, the
effects of the transverse expansion are generally more dramatic.
The rarefaction wave reaches the center after about 12 fm/$c$, and
all of the matter participates in the transverse flow at the moment
of hadronization.  The more rapid cooling of the system is also 
indicated by the decreasing separation of the contours at later times.  
This affects the chemical equilibration process in several competing ways.

We note that,  as before, the quark fugacities lag behind the gluon 
fugacities at all times and all radial distances. This is not surprising, 
considering the lower starting value for $\lambda_q$.  
We also see again that the fugacities in the interior, at first,
increase with passage of time. However, beyond $N=31$, corresponding
to about 8 fm/$c$ near the center, when the transverse flow has been
substantially developed, the fugacities begin decreasing again, except 
at $r=0$, where they continue to increase, as all the radial derivatives 
in (\ref{master}) are small there.

Let us pause to understand this unexpected feature.  According to 
(\ref{master1}), the evolution of the parton densities is 
governed by $\partial_\mu (n u^\mu)$ which consists of
two terms:  $u^\mu \partial_\mu n$ gives the rate of change of 
the density in the comoving frame, and $n\, \partial_\mu u^\mu$ 
describes the rate of change of the density due to the expansion of 
the fluid element in the comoving frame \cite{str}.  Once the 
transverse expansion of the fluid  starts developing, the second term 
and also the radial derivatives in (\ref{master})
grow very rapidly and drive the system away from chemical equilibrium. 

We thus see that two mechanisms potentially contribute to the depletion 
of the parton densities: expansion and flow.  Let us concentrate on the
consequences of the latter, as the consequences of the expansion alone are
easy to visualize.  Figure 4 depicts a fluid element at a distance $r$ 
from the axis at some instant of time, in a central collision
for the conditions of: (a) no flow, (b) uniform transverse flow, and 
(c) a differential transverse flow with the velocity field $v_r(r,t)$ 
as obtained in the present work.
When there is no flow, the thermal motion of the partons brings the same
number of partons into the volume element as are leaving it on all sides. 
Thus any production of new partons due to the chemical reactions in the
volume element will increase the fugacities, which will hence rise
monotonically since the right hand sides of the master equations
(\ref{master_long},\ref{master}) remain positive \cite{biro,new}.

\begin{figure}
\epsfxsize=3.25in
\epsfbox{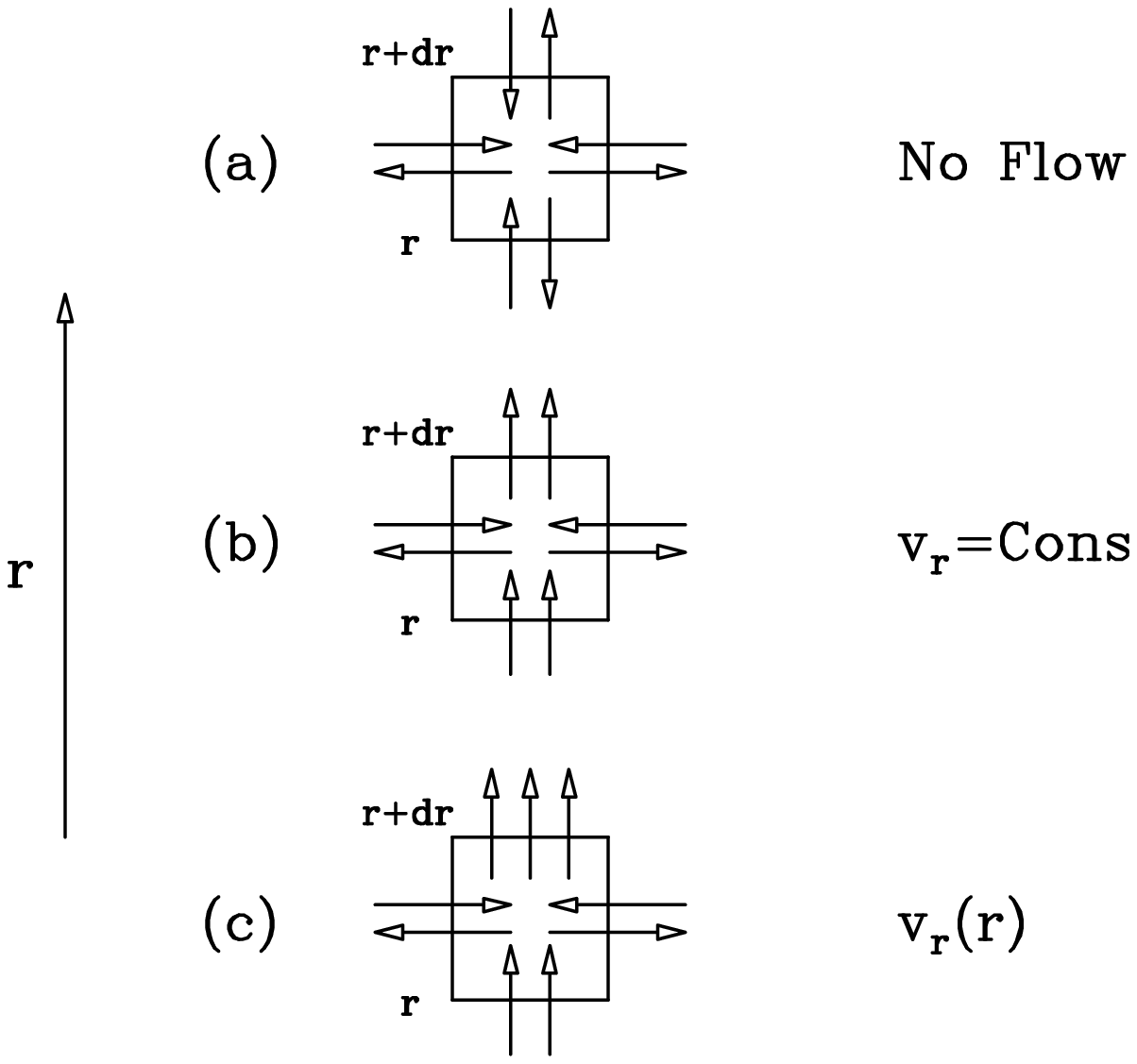}
\caption{ Depletion of partons for, (a) no flow, (b) uniform flow,
and (c) flow with velocity gradient in a fluid element at $r$.}
\end{figure}

Next consider the condition of a uniform transverse flow, which is
sometimes used as a simplifying assumption \cite{kampfer}.
This scenario still does not produce a net out-flow of partons from 
the volume element, because the loss of partons on one side is exactly
balanced by the gain of partons flowing in from the opposite side.
Thus, if there is a creation of new partons due to chemical reactions 
then, as before, the fugacities will increase with time.

However, when the transverse velocity is driven by the pressure gradient
as in our calculation, $v_r(r+dr) > v_r(r)$. Then the velocity gradient will
ensure that more partons are leaving the fluid element at $r+dr$ than are
entering at $r$. This leads to an additional depletion of the parton
density, beyond the dilution caused by the overall expansion.  If the 
velocity gradient is sufficiently large, the production of new partons by
the chemical reactions cannot cope with the depletion and 
the fugacities will start to fall. The richness of this scenario is 
completely absent when the simplifying assumptions of no-flow or of
a uniform transverse flow are made.

Let us return to the discussion of the consequences of the transverse
flow on the momentum distribution of the partons at LHC energies.
In Figs.~5a--b, we show the evolution of the gluon and quark 
momentum distributions over the course of the expansion. 
We see that the distribution at later times is strongly affected by 
the flow. It is again apparent that there is an approximate balance
between the decreasing temperature and the increasing transverse velocity
which keeps the parton distributions at high $p_T$ largely independent 
of $N$.  This is reminiscent of a similar behavior observed in pion 
spectra when transverse flow is introduced \cite{cape}.

\begin{figure}
\epsfxsize=3.25in
\epsfbox{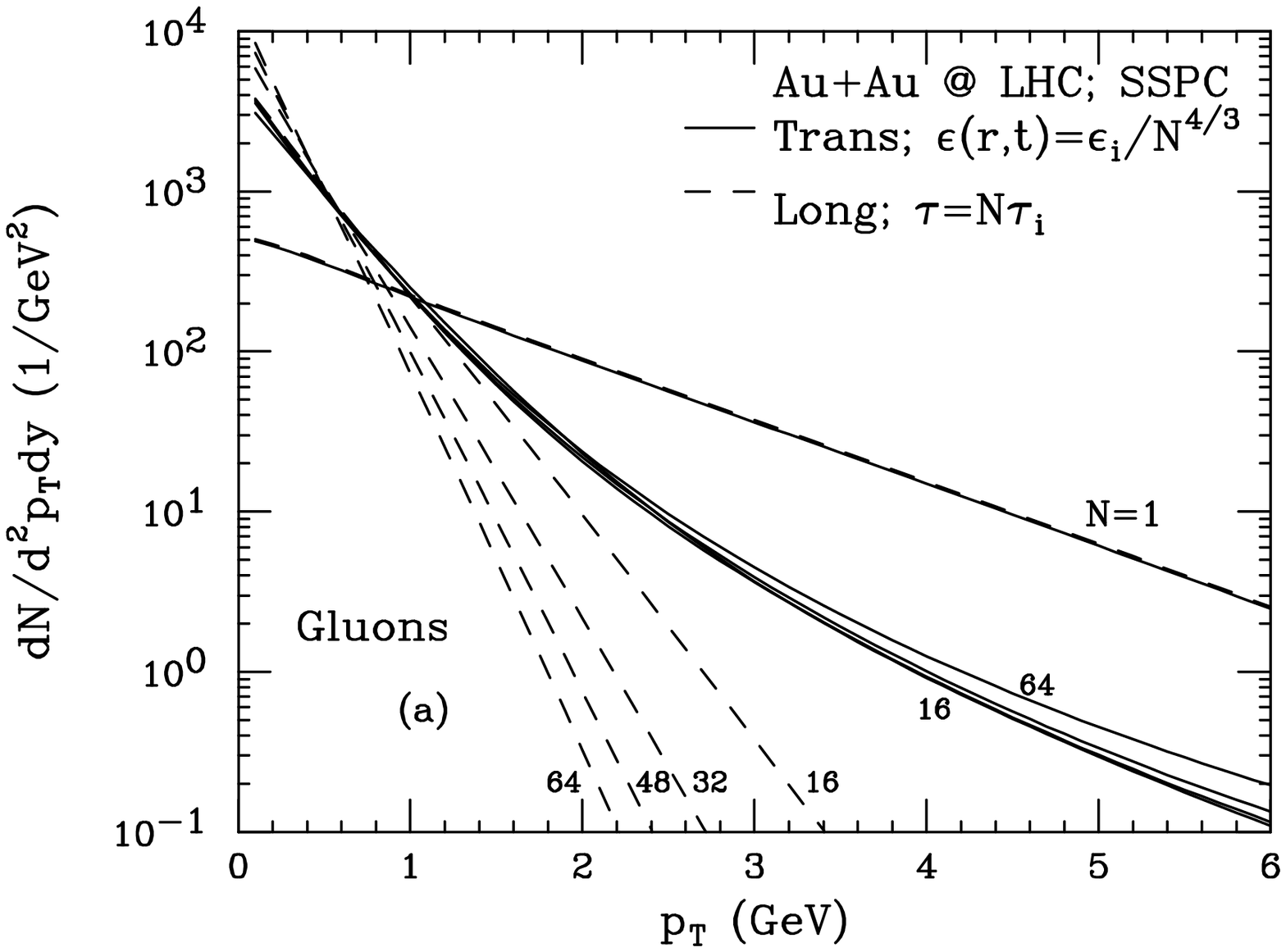}
\epsfxsize=3.25in
\epsfbox{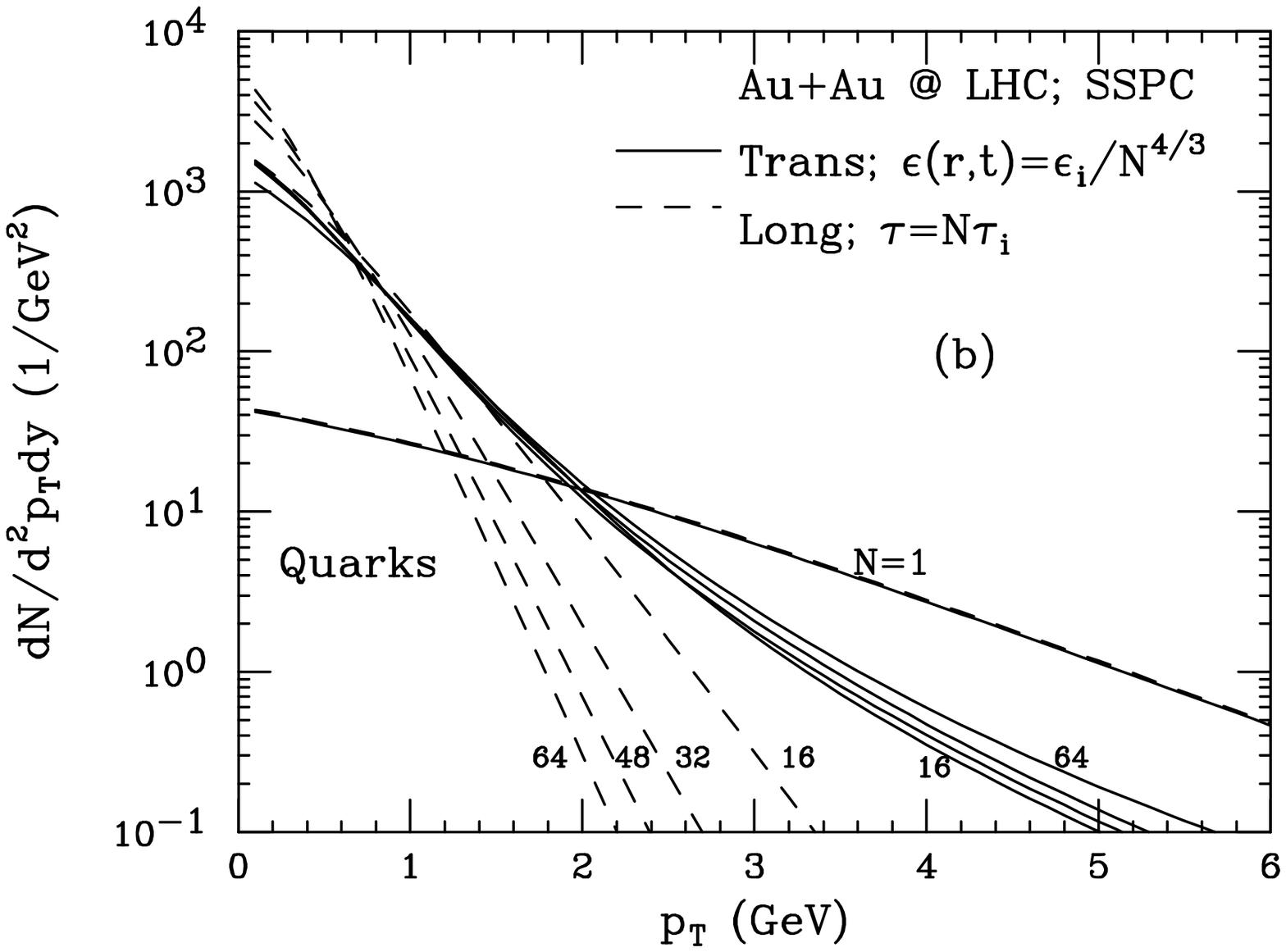}
\epsfxsize=3.25in
\epsfbox{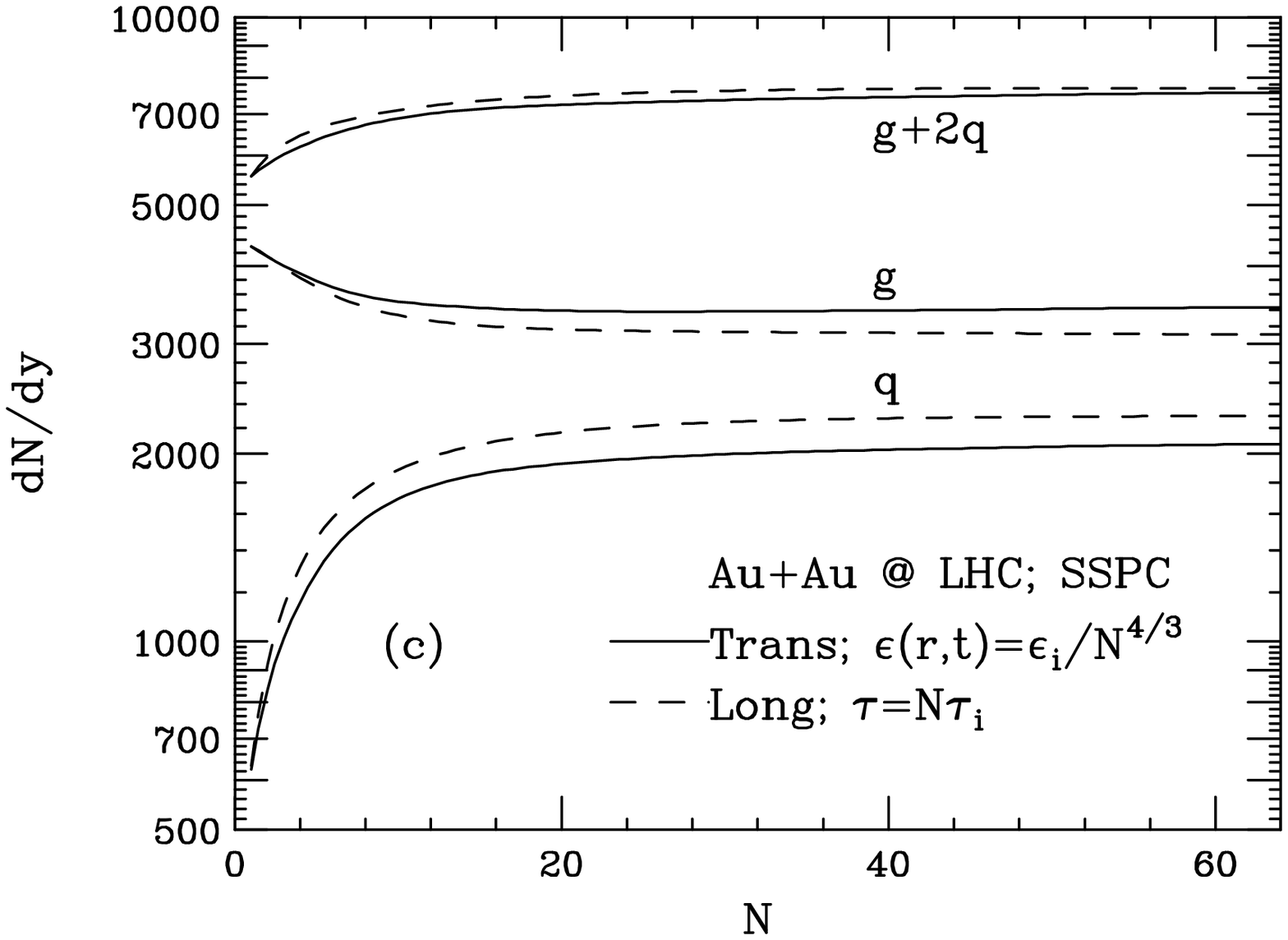}
\caption{ (a)--(c) Same as Fig.~2, for LHC.}
\end{figure}

Figure 5c shows our results for the evolution of the parton multiplicity. 
We see a significant production of quarks and antiquarks, in fact 
an increase by a factor of about 2.5, and a decrease in the number 
of gluons by about 25\%, as a result of the chemical reactions in the 
plasma. This results in a net increase of the parton multiplicity 
by approximately 40\%.  As before at RHIC energies, we find that the 
transverse flow does not strongly affect the net production of partons, 
although it influences the individual abundances of gluons and quarks.

It is probably difficult to probe the off-chemical equilibrium nature 
of the light quark and gluon fractions in the plasma.  However, we can
think of a practical way of testing the chemical equilibrium in the
cases of heavier quark flavors, viz., strange, charm, and even bottom
quarks, by inferring this information from the ratios of the (valence) 
quark content of the emitted hadrons \cite{cleymans,johanna}. We can
analogously define the ratio of heavier quarks  to light quarks as
\begin{equation}
{\cal R}=\frac{\langle Q + {\bar Q} \rangle}{0.5\left[\langle u+\bar{u} 
\rangle +\langle d +\bar{d}\rangle \right]}
\end{equation}
where $Q$ stands for one of the heavy quarks. It is not unlikely that
this ratio can be largely determined from the final production of
strange, charm, or bottom mesons and pions\cite{chem}.

Simulations of the pure SU(3) lattice gauge theory have shown that
the speed of sound drops significantly as the temperature approaches
the phase transition temperature from above \cite{boyd}. If this
observation carries over to full QCD, our results could be modified.
A lower speed of sound generally leads to a reduced production of
transverse flow and, hence, to an increased life-time of the QGP.
This effect should be especially pronounced at RHIC energies \cite{RG95}.

Can we devise means for identifying the large transverse flow velocities
predicted here?  As mentioned earlier, the life-time of the mixed 
phase, which follows the QGP phase, gets shorter if the transverse 
velocity increases.  Because the transverse flow will remain frozen
in the mixed phase, the final hadronic flow pattern should be a 
rather reliable measure of the transverse velocity established during
the QGP phase.  In a forthcoming publication, we shall report on the
effect of the transverse flow on the $p_T$-spectra of charmed hadrons
which are considered to be good probes of flow due to their large mass.
In the next section we look at the spectra of single photons and lepton 
pairs and see how far could they be affected by these consideration.
 
\subsection{Other Initial Conditions}

Let us now briefly examine the consequences of using different initial 
thermalization conditions.  We only present the results for the hadronization
contour, $\epsilon(r,t)=1.45$ GeV/fm$^3$, and for the gluon and
quark fugacities attained along this contour for RHIC (Fig.~6a--b)
and LHC (Fig.~7a--b) energies.  We select several initial conditions 
obtained by modifying predictions of the HIJING model, as listed in Table I.
The original prediction of the HIJING model \cite{hijing} is denoted
as I.  The second set (II) of initial conditions is obtained by increasing
the initial fugacities, and hence the energy density, by a factor four. 
The third set (III) is obtained by retaining the fugacities of the second 
set, but decreasing the temperature to account for the production of
soft partons from the color field, by some arbitrary amount.  
It is of interest to recall that the Parton Cascade Model of Geiger \cite{kg}
predicts an energy density of about 120 GeV/fm$^3$, and a temperature of
about 590 MeV, at $\tau=$ 0.25 fm/$c$ in central collisions of two
gold nuclei at  RHIC energies. 

\begin{figure}
\epsfxsize=3.25in
\epsfbox{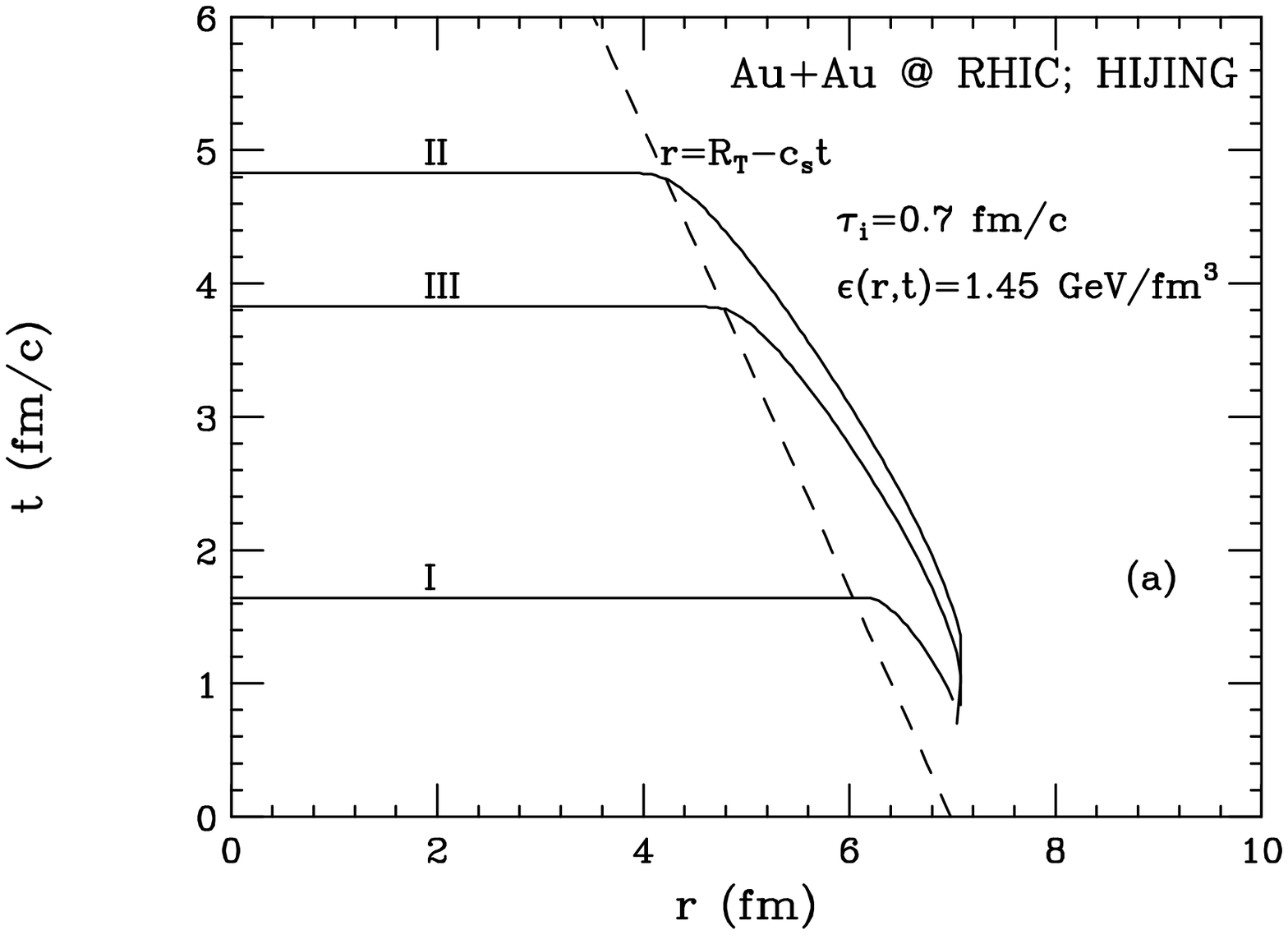}
\epsfxsize=3.25in
\epsfbox{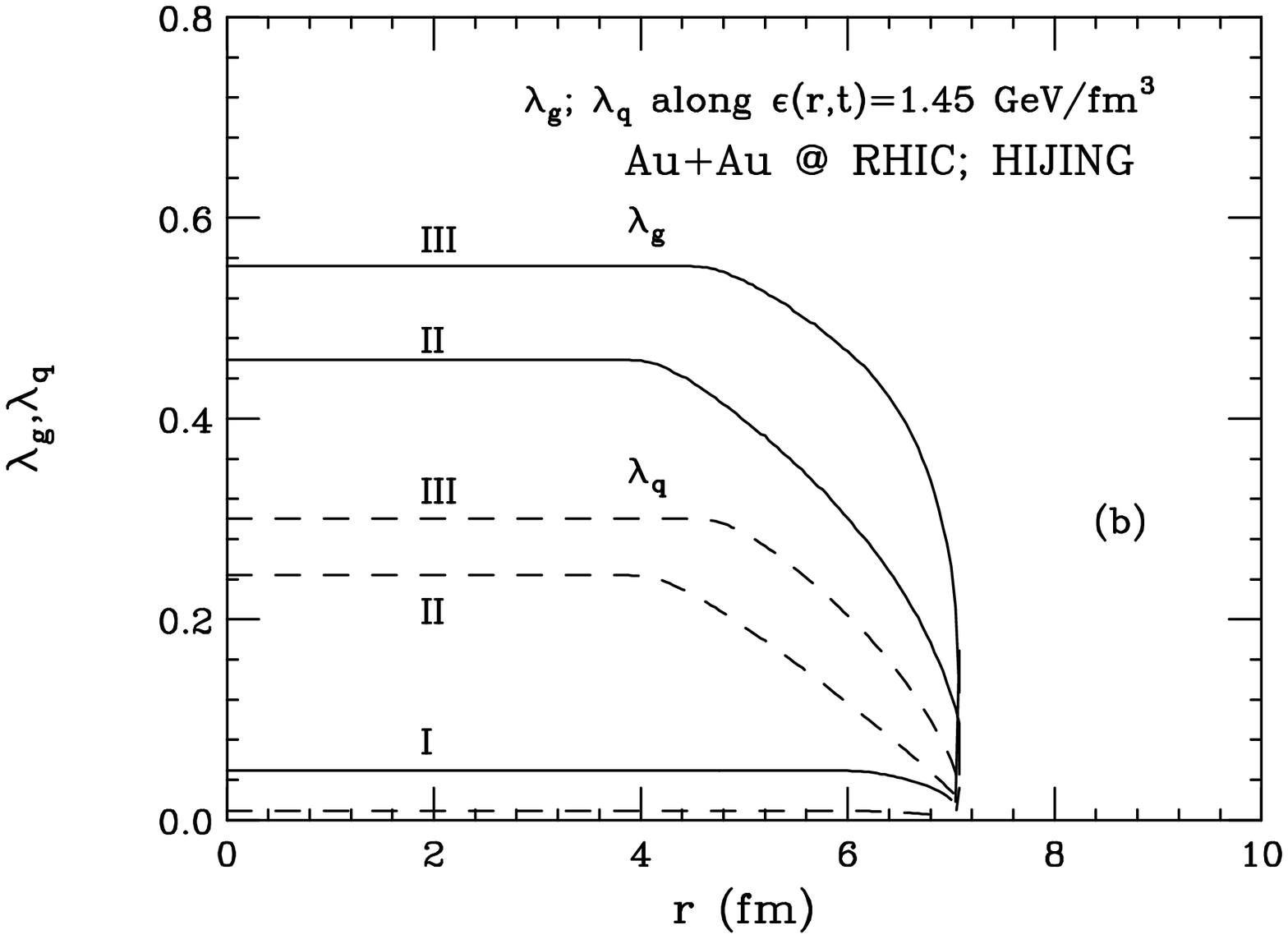}
\caption{ (a) Constant energy density contours for $\epsilon=$ 1.45 GeV/fm$^3$
for the initial conditions obtained from HIJING, given in Table I.
(b) Gluon and quark fugacities along these contours.}
\end{figure}

\begin{figure}
\epsfxsize=3.25in
\epsfbox{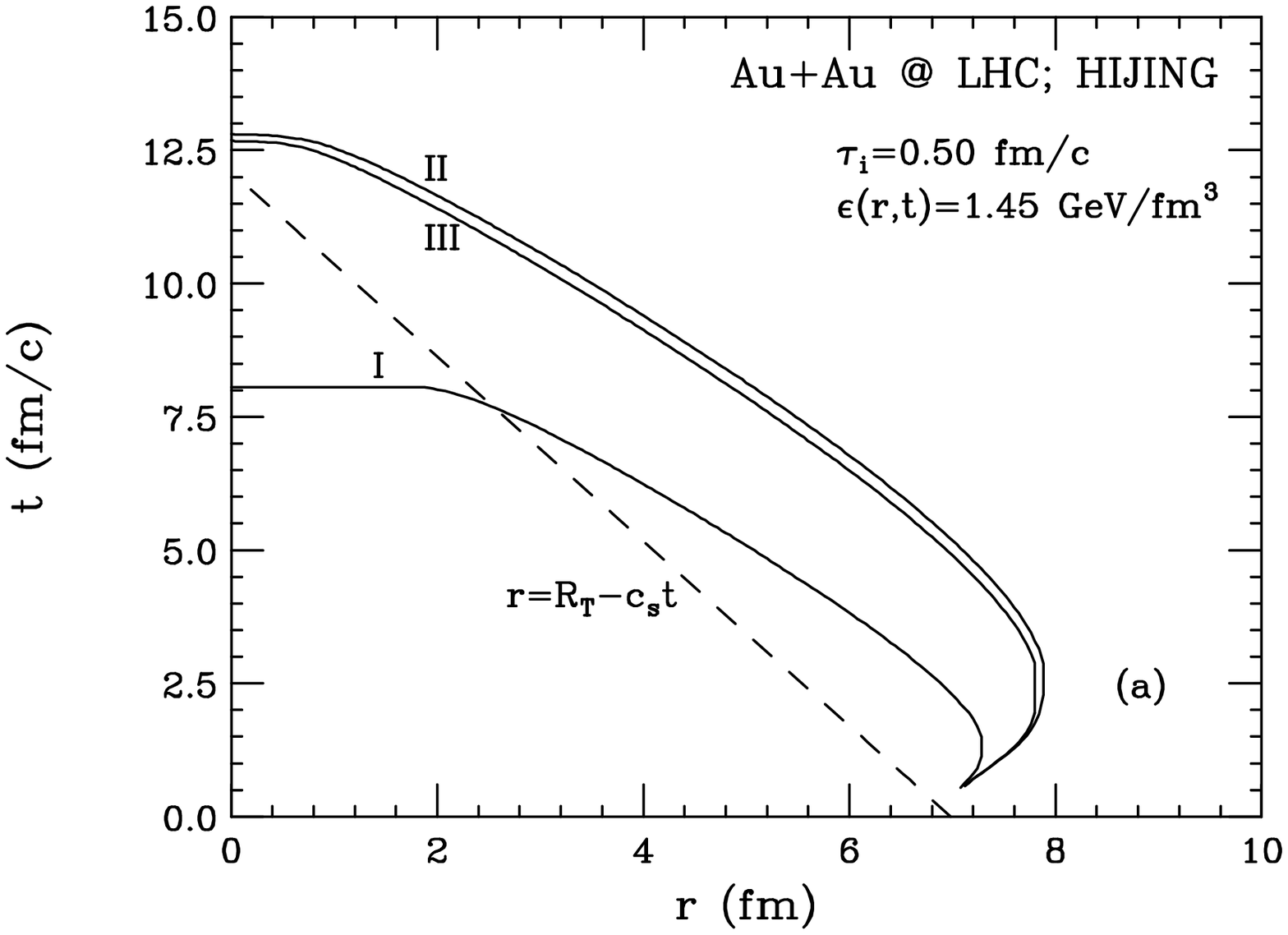}
\epsfxsize=3.25in
\epsfbox{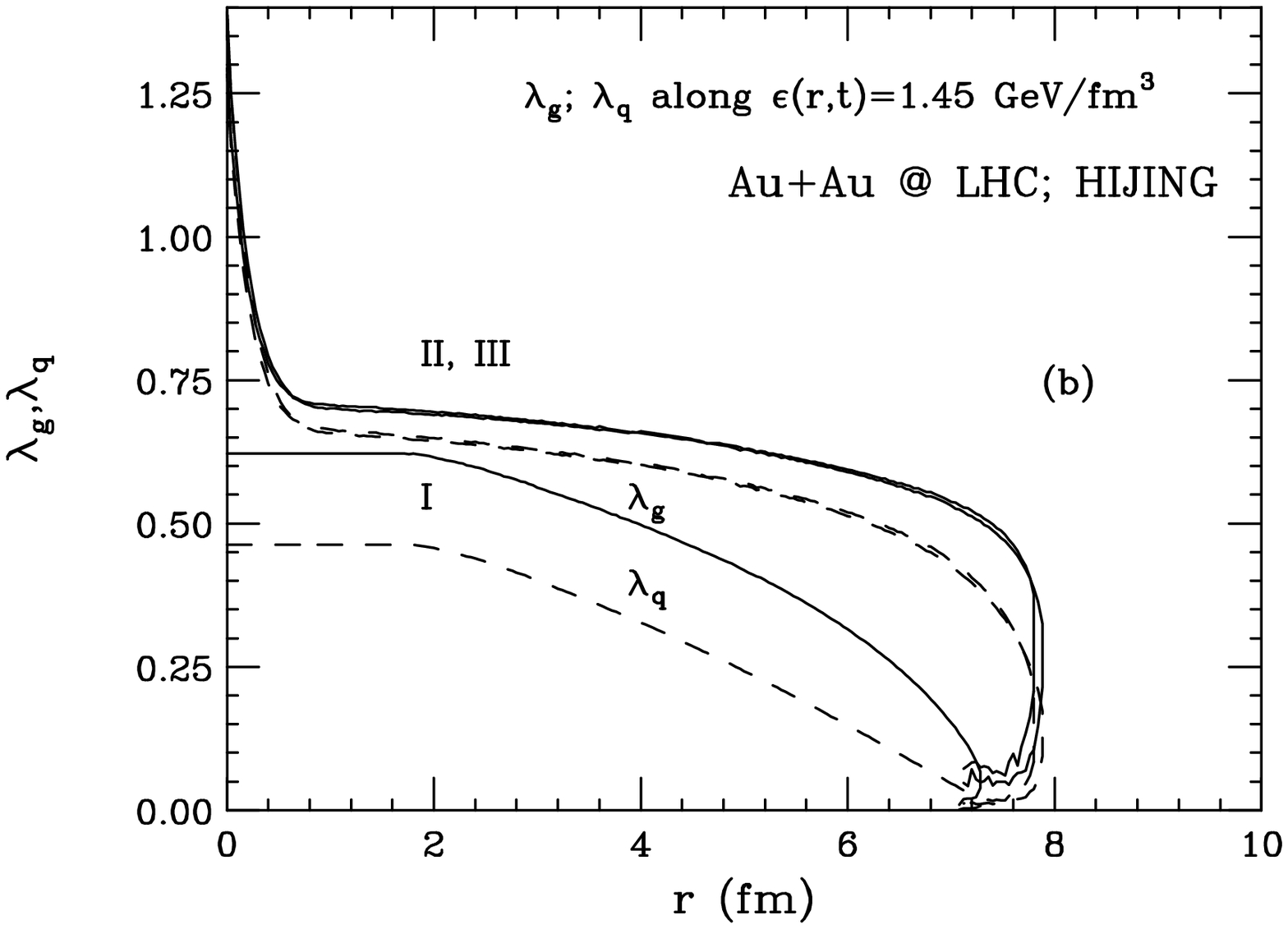}
\caption{ Same as Fig.~6, for LHC.}
\end{figure}

If we use the original predictions (I) 
of the HIJING model, then the life-time of the QGP phase is very small 
(less than 2 fm/$c$), and the matter at the moment of hadronization is 
extremely dilute compared with chemically equilibrated matter. Changing
the initial fugacities by a factor of 4 increases the life-time of the QGP
phase considerably and also brings the system close to chemical 
equilibrium. Note that the initial fugacities differ by one order of 
magnitude between the conditions I and III at RHIC energies. Note also that
the final fugacities at large $r$ are quite different, even though the
energy densities are identical, signaling very different temperatures
at the end of the QGP phase. 

Results for LHC energies indicate the effect of transverse expansion for $r
\ge 4$ fm, for the original predictions of HIJING, marked as I, and the entire
fluid is affected by the flow for the other initial conditions, which assume
much larger initial energy densities and fugacities. The final fugacities 
for the conditions II and III are essentially identical and show the effect of 
evolution away from chemical equilibrium, as the flow velocity gets large. 
The transverse dimension of the system grows by almost 1 fm during the 
QGP phase itself for the initial conditions II and III.  Since the 
fugacities shown in these two figures are for the same energy density
$\epsilon_f$, we note that the final temperature for the standard case 
(I) will be much larger than in the scenarios II and III.

\section{\bf THERMAL PHOTONS AND LEPTON PAIRS}
\subsection{Photon Spectra}

Thermal photons and lepton pairs are primary probes of all the stages of 
the nuclear reaction.  They are also expected to carry valuable imprints
of the transverse expansion of the system. Thermal photons from the
quark-gluon plasma have their origin in the 
Compton ($qg\rightarrow q\gamma$)  
and annihilation processes ($q\bar{q}\rightarrow g\gamma$). 
It is useful to recall the expressions for the rates for these 
contributions for a non-equilibrated quark-gluon plasma \cite{strickland}.
For the approximations of the fugacities that we have used, the Compton
process contributes with the rate
\begin{eqnarray}
E\frac{dN_\gamma}{d^3p\,d^4x}= \frac{2\alpha\alpha_s}{\pi^4} 
\lambda_q \lambda_g T^2 & &e_q^2 \exp(-E/T)\nonumber\\
& &\left[ \ln\left(\frac{4ET}{k_c^2}\right)
+\frac{1}{2}-C\right],
\label{comp}
\end{eqnarray}
and the rate of the radiative annihilation process is,
\begin{eqnarray}
E\frac{dN_\gamma}{d^3p\,d^4x}= \frac{2\alpha\alpha_s}{\pi^4}
 \lambda_q \lambda_{\bar{q}}
T^2& & e_q^2\exp(-E/T)\nonumber\\
& &\left[ \ln\left(\frac{4ET}{k_c^2}\right)
-1-C\right].
\label{ann}
\end{eqnarray}
Here $C=$ 0.577721\ldots, $e_q$ is the electric charge of the quark and
the parameter $k_c$  is related to the thermal mass of the quarks in the 
medium.

Recalling that our plasma is gluon-rich and quark-poor, we anticipate
that the yield of the thermal photons would be dominated by the 
Compton contribution. In Figs.~8 and 9, we give our results for RHIC
and LHC energies, respectively, for the initial conditions obtained from 
the SSPC \cite{sspc}.  We show the Compton and annihilation contributions 
separately for the scenario involving transverse expansion. The results
for longitudinal expansion are given for comparison.

\begin{figure}
\epsfxsize=3.25in
\epsfbox{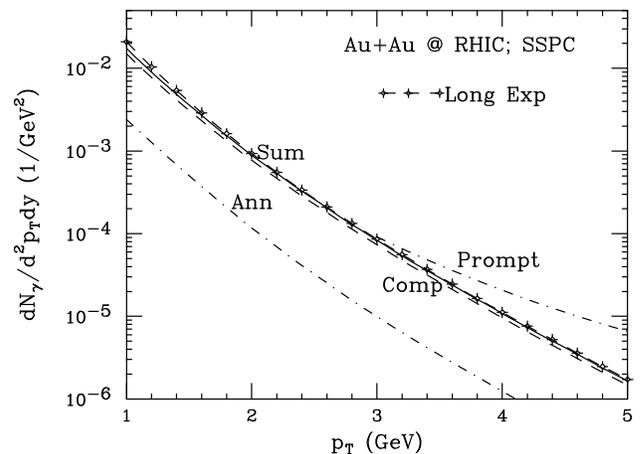}
\caption{ Distribution of thermal photons from the QGP phase at RHIC.
The Compton and annihilation yields, and their sum are shown for 
the case with transverse expansion.
Results are also given for a purely longitudinal flow. Prompt photons,
whose production is governed by structure functions, are seen to dominate 
the yield for $p_T >$ 3--4 GeV.}
\end{figure}

\begin{figure}
\epsfxsize=3.25in
\epsfbox{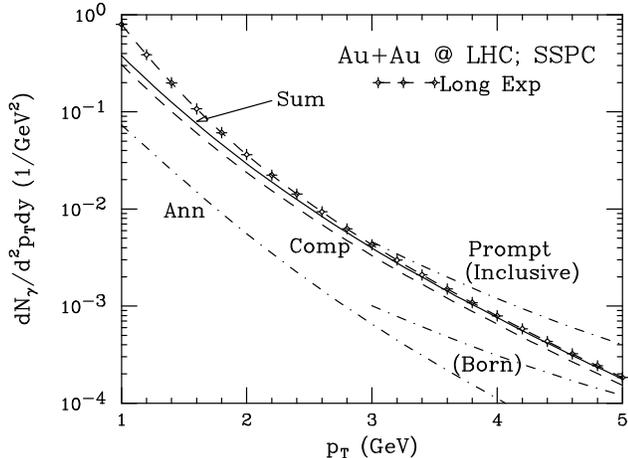}
\caption{ Same as Fig.~8 at LHC. The prompt photons from fragmentation
of quark jets are also shown.}
\end{figure}

We see that the transverse expansion of the fluid leaves the thermal photon
production  from the QGP at RHIC energies essentially unchanged for
$p_T>1$ GeV. The  reduction of the thermal photon production at lower $p_T$
will be difficult to identify, as it will be overshadowed by photons from
hadronic reactions and hadronic decays. We also see that prompt photons
\cite{jean} will dominate the yield beyond $p_T\ge 4$ GeV. At first
this result looks surprising, in view of the large modifications
of the final parton spectra seen earlier (cf.~ Figs.~2 and 5). 
However, note that the rates for photon emission (\ref{comp},\ref{ann})
carry an additional weight factor  of $T^2$, which suppresses the 
contributions from later times.  Note also that the time available
for radiation at a given temperature is significantly reduced by the 
transverse expansion.  Overall, we conclude that the thermal photon
production from the plasma phase is remarkably insensitive to the
transverse expansion at RHIC energies.

The production of thermal photons with large $p_T$ remains essentially
unchanged at LHC, as well, as they have origin in the early hot stages, when
the flow effects are still small.  On the other hand, the production of 
low $p_T$ photons decreases considerably, due to reduction in the space-time 
volume occupied by colder matter in the presence of transverse expansion. 
While the prompt photon production \cite{jean} may remain lower than the 
thermal photon yield up to $p_T=5$ GeV, the background contribution
of photons fragmented off high-$p_T$ quark jets is large.

\subsection{Lepton Pair Spectra}

The mass distribution of lepton pairs obtained from quark annihilation are
given in Figs.~10 and 11, for RHIC and LHC energies, respectively, without 
and with transverse expansion. The Drell-Yan contribution \cite{sean} is also
shown. We again see a negligible effect of the flow on the mass distribution 
of lepton pairs at RHIC energies, but a factor of 2--3 reduction is seen in 
the yield at LHC energies at invariant masses of 1--2 GeV, due to the reduced
life-time of the plasma. We also note that the Drell-Yan contribution
dominates the yield beyond $M\ge 4$ GeV at LHC, whereas this zone shifts
to much smaller $M$ at RHIC.  This is primarily due to very low quark
densities of the plasma.

\begin{figure}
\epsfxsize=3.25in
\epsfbox{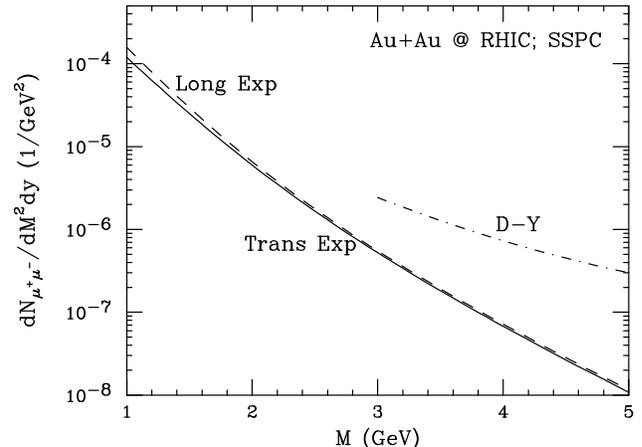}
\caption{ Mass distribution of thermal lepton pairs from the QGP phase at RHIC
with (solid curves) and without (dashed curves) transverse flow. The Drell-
Yan contribution is also given.}
\end{figure}

\begin{figure}
\epsfxsize=3.25in
\epsfbox{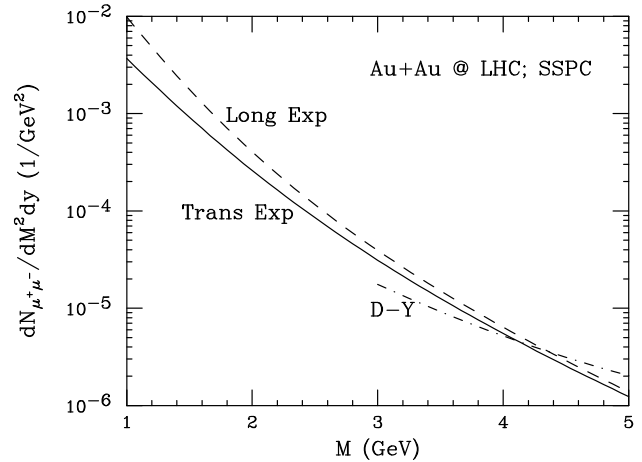}
\caption{ Same as Fig.~10 for LHC.}
\end{figure}

These small deviations in the production of thermal photon and lepton
pairs due to the transverse expansion are rather astounding.  We have 
already noted that this has its origin in the strong temperature dependence
of the production rates, combined with the accelerated cooling of a
transversely expanding plasma.  This raises the question whether it is
at all possible to identify the presence of transverse flow from the
lepton pair spectra.  Recall that the transverse mass distribution of the 
lepton pairs from the QGP in presence of a flow will be given by,
\begin{eqnarray}
\frac{dN_{\mu^+\mu^-}}{dM^2\,d^2M_T\,dy}=\frac{\alpha^2}{2\pi^3}
& &\lambda_{q}\lambda_{\bar{q}}
e_q^2 \int \tau\,d\tau\,rdr\,\nonumber\\
& &I_0\left(\frac{\gamma v_r p_T}{T}\right)
K_0\left(\frac{\gamma M_T}{T}\right).
\end{eqnarray}
In the absence of a flow, this reduces to
\begin{equation}
\frac{dN{\mu^+\mu^-}}{dM^2\,d^2M_T\,dy}=\frac{\alpha^2}{4\pi^3}
\lambda_{q}\lambda_{\bar{q}}
e_q^2 R_T^2 \int \tau\,d\tau
K_0\left(\frac{M_T}{T}\right),
\end{equation}
which scales with $M_T$ in a characteristic manner which is known as 
$M_T$- scaling \cite{asakawa,dks}. This scaling will be violated in the
presence of a transverse flow \cite{dks}.

In Figs.~12 and 13 we present our predictions for the ratios of the 
transverse mass distributions for $M=1$ GeV to that for $M=$ 2, 3, and 4
GeV  at RHIC and LHC energies, respectively. As indicated earlier these
ratios are identically one for a purely longitudinal expansion.  The 
deviations predicted here are not large, but we hope that they will be
observable in the high statistics data that will be forthcoming from 
RHIC and LHC.  Additional information about the parton fugacities is
contained in the absolute yields of the photon and lepton pair spectra,
because the thermal photon yield is dominated by the Compton processes
($\propto \lambda_g \lambda_q$), whereas the lepton pair yield is mainly
due to quark-antiquark annihilation ($\propto \lambda_q^2$).

\begin{figure}
\epsfxsize=3.25in
\epsfbox{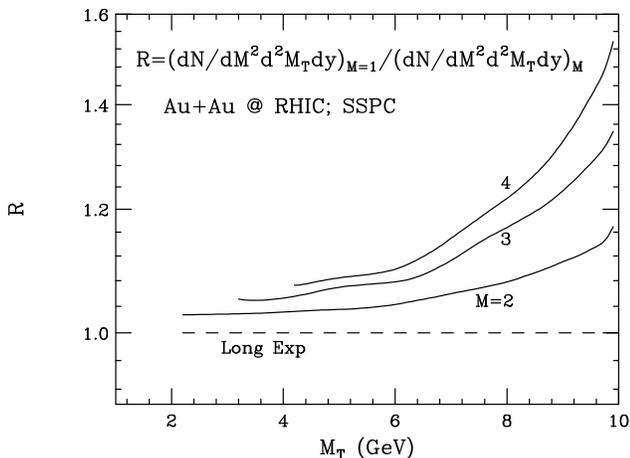}
\caption{ Ratio of transverse mass distribution for thermal
lepton pairs for $M=$ 1 GeV to that for a given $M$.
In absence of flow, it is identically unity.}
\end{figure}

\begin{figure}
\epsfxsize=3.25in
\epsfbox{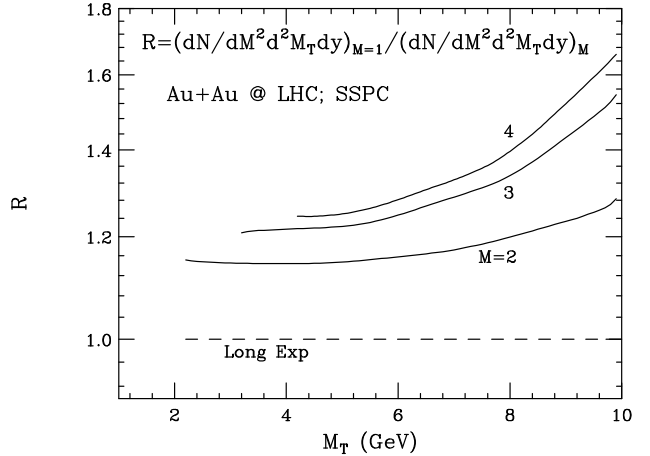}
\caption{ Same as Fig.~12 at LHC.}
\end{figure}

\section{\bf SUMMARY}
We have studied the evolution of an equilibrating and transversely
expanding quark-gluon plasma, with initial conditions that have been 
obtained either from the self-screened parton cascade model or from 
the HIJING model. We have compared the results from a boost invariant 
longitudinal, as well as from a transverse expansion of the plasma. 
The velocity gradients generated due to flow were found to affect the 
individual concentrations of the quarks, antiquarks, and gluons, without 
affecting the total final entropy. 

Large transverse velocities are developing by the end of the QGP 
phase, especially at LHC energies. The flow is also seen to drive the
system away from chemical equilibrium. As the process of chemical 
equilibration accelerates cooling, and as the rates of emission of
photons and lepton pairs depend strongly on the temperature, the final 
spectra are only slightly affected by them. This has its origin in 
the fact that photons and lepton pairs originate from the early hot stage
when the flow is still small. However, the transverse flow is seen to 
introduce a slight violation of the so-called $M_T$ scaling of
the lepton pair spectra.  Likely consequences on the post-QGP epoch 
could be a copious production 
of low $p_T$ pions from the fragmentation of residual gluons. 
We also note that the temperature of the matter at the end of the 
so-called QGP phase, may be higher at the surface than in the center
of the reaction volume.

\section*{\bf Acknowledgments} 
We acknowledge useful comments from  Pradip Kumar Roy, Sourav Sarkar, 
and Bikash Sinha. This work was supported in part by a grant from the 
U.S. Department of Energy (DE-FG02-96ER40945).

\newpage

\begin{table}
\caption{Initial conditions for the hydrodynamical expansion phase
in central collision of two gold nuclei
at BNL RHIC and CERN LHC energies from SSPC and HIJING models.}
\begin{center}
\begin{tabular}{|l|c|c|c|c|c|} 
\hline
& & & &&\\
Energy &
$\tau_i$ & $T_i$ & $\lambda_g^{(i)}$&
$\lambda_q^{(i)}$ & $\epsilon_i$ \\
& & & &&\\
 & (fm/$c$) & (GeV)&-&-& (GeV/fm$^3$) \\
& & & & &\\ \hline
{\bf SSPC} & & & &&\\
&&&&&\\ 
 RHIC & 0.25 & 0.668 & 0.34 & 0.064 & 61.4  \\
& & & & &\\ 
LHC & 0.25 & 1.02 & 0.43 & 0.082 & 425 \\
& & & & &\\ 
{\bf HIJING} &&&&&\\
&&&&&\\
 RHIC, I & 0.7 & 0.55 & 0.05 & 0.008& 4.0 \\
& & & & &\\ 
 RHIC, II & 0.7 & 0.55 & 0.20 & 0.032&15.8 \\
& & & & &\\
 RHIC, III & 0.7 & 0.40 & 0.53 & 0.083& 11.7\\
& & & & &\\ 
 LHC, I & 0.5 & 0.82 & 0.124 & 0.02&48.6 \\
& & & & &\\ 
 LHC, II & 0.5 & 0.82 & 0.496 & 0.08& 194 \\
& & & & &\\
 LHC, III & 0.5 & 0.72 & 0.761 & 0.118&176 \\
& & & & &\\ 
\hline
\end{tabular}
\end{center}
\end{table}

\end{document}